\documentclass[final,onefignum,onetabnum]{atemplate}

\usepackage{lipsum}
\usepackage{amsfonts}
\usepackage{graphicx}
\usepackage{epstopdf}
\usepackage{algorithmic}
\usepackage{longtable}
\usepackage{amsmath}
\usepackage{pgfplots}
\usepackage{csvsimple}
\usepackage{tikz}
\usepackage{subfigure}
\usepackage{amssymb} 
\usepackage[utf8]{inputenc}
\usepackage{color}
\usepackage{url}
\usepackage{mathtools}
\usepackage{xcolor}
\usepackage{colortbl}
\usepackage{booktabs}
\usepackage{multirow}
\usepgfplotslibrary{groupplots}
\usepgfplotslibrary{fillbetween}
\ifpdf
  \DeclareGraphicsExtensions{.eps,.pdf,.png,.jpg}
\else
  \DeclareGraphicsExtensions{.eps}
\fi

\newcommand{\nat}{\mathbb{N}}
\newcommand{\real}{\mathbb{R}}
\newcommand{\ex}{\mathbb{E}}
\newcommand{\pr}{\mathbb{P}}
\newcommand{\siga}{\mathcal{F}_{k - \lambda}}
\newcommand{\bound}{M_{k-\lambda}}
\newtempremark{remark}{Remark}
\newtempremark{hypothesis}{Hypothesis}
\crefname{hypothesis}{Hypothesis}{Hypotheses}
\newtempthm{claim}{Claim}
\newtempthm{assumption}{Assumption}
\crefname{assumption}{Assumption}{Assumptions}

\definecolor{white}{HTML}{FFFFFF}
\definecolor{red1}{HTML}{FF9292}
\definecolor{red2}{HTML}{FF6161}
\definecolor{red3}{HTML}{FF3434}
\definecolor{red4}{HTML}{FD0000}

\headers{Tracking and Stopping for Least Squares}{Pritchard \& Patel}

\title{Towards Practical Large-scale Randomized Iterative Least Squares Solvers through Uncertainty Quantification\thanks{Submitted to the editors August 9, 2022.
\funding{Authors are supported by UW-Madison WARF Award AAD5914.}}}

\author{Nathaniel Pritchard\thanks{Department of Statistics, University of Wisconsin - Madison
  (\email{npritchard@wisc.edu}).}
\and Vivak Patel\thanks{Department of Statistics, University of Wisconsin - Madison
  (\email{vivak.patel@wisc.edu},\url{vivakpatel.org}).}
}

\usepackage{amsopn}

\DeclareMathOperator{\col}{col}
\DeclareMathOperator{\rank}{rank}
\DeclareMathOperator{\argmin}{argmin}

\ifpdf
\hypersetup{
  pdftitle={Gradient Tracking and Stopping for Generalized Coordinate Descent},
  pdfauthor={N. Pritchard and V. Patel}
}
\fi

\allowdisplaybreaks
\begin{document}

\maketitle

\begin{abstract}
As the scale of problems and data used for experimental design, signal processing and data assimilation grow, the oft-occuring least squares subproblems are correspondingly growing in size. As the scale of these least squares problems creates prohibitive memory movement costs for the usual incremental QR and Krylov-based algorithms, randomized least squares problems are garnering more attention. However, these randomized least squares solvers are difficult to integrate application algorithms as their uncertainty limits practical tracking of algorithmic progress and reliable stopping. Accordingly, in this work, we develop theoretically-rigorous, practical tools for quantifying the uncertainty of an important class of iterative randomized least squares algorithms, which we then use to track algorithmic progress and create a stopping condition. We demonstrate the effectiveness of our algorithm by solving a 0.78 TB least squares subproblem from the inner loop of incremental 4D-Var using only 195 MB of memory.
\end{abstract}

\begin{keywords}
  random sketching, linear systems, iterative methods, residual estimation, stopping criterion, least-squares, Coordinate Descent  
\end{keywords}

\begin{AMS}
  65F10, 65F25, 60F10, 62L12
\end{AMS}

\section{Introduction}\label{sec:Intro}
Least squares problems are regularly solved as core subproblems in a variety of important algorithms for experimental design \cite{krishnan2011partial,chen2003sparse}, signal processing \cite{selesnick2013least,so2011linear}, data assimilation  \cite{talagrand1987variational,gurol:hal-03224132}, and uncertainty quantification \cite{Tarakanov_2019,singh}. Moreover, these least squares subproblems are growing in both the number of equations and the dimension of the unknown variables owing to two pressures: (1) improvements in technology have increased the permeation of higher-frequency sensors, which grows the volume of data being used and which, in turn, (usually) increases the number of equations in the least squares subproblem; and (2) the growing desire for more accurately simulating models (e.g., using finer meshes for partial differential equation models) increases the number of unknown variables in the least squares problems. 

Unfortunately, the growth of least squares subproblems is a challenge for commonly used solvers. For instance, solving a least squares problem with many observations can be addressed in a memory-efficient manner using an incremental QR algorithm \cite{MillerGentl}, so long as the resulting upper triangular term can be fit in memory. Unfortunately, if the number of unknowns is sufficiently larger, this least squares incremental QR algorithm will be unable to store and manipulate the resulting upper triangular matrix without substantial slowdowns induced by memory movement costs. As another example, Krylov-based least squares solvers can also be efficiently deployed \cite{hayami}, so long as matrix-vector and matrix-transpose-vector products can be efficiently computed. Unfortunately, if the system is sufficiently large that it cannot be stored in memory, then Krylov-based least squares solvers are substantially slowed down also because of the memory movement costs needed to read in the matrix multiple times per iteration \cite{Lam1991TheCP}.

As these challenges to standard solvers are driven by size, randomized least squares solvers (e.g., iterative Hessian sketch \cite{pilanciIHS} and generalized column subspace descent \cite{patel2021implicit,patel2021convergence,patel2022randomized}) seem to be promising alternatives as they are able to compress the information in the original linear system to more manageable dimensions. However, such iterative randomized least squares solvers must first overcome a key practical challenge: as such solvers would be called repeatedly within an iterative algorithm, their solution accuracy must be controlled so as to ensure algorithmic efficiency. For example, in incremental 4D-Var \cite{courtier1994strategy}, a least squares subproblem  occurs at every iteration of the algorithm. Indeed, in the initial few iterations, the least squares subproblem only needs to be solved to low accuracy as this is usually enough to generate progress quickly, while later iterations will demand that the subproblem be solved to higher accuracy. Thus, achieving such control over the least squares subproblem solver's accuracy ensures the efficiency of the overall algorithm.

\begin{figure}[ht]
    \centering
    \input{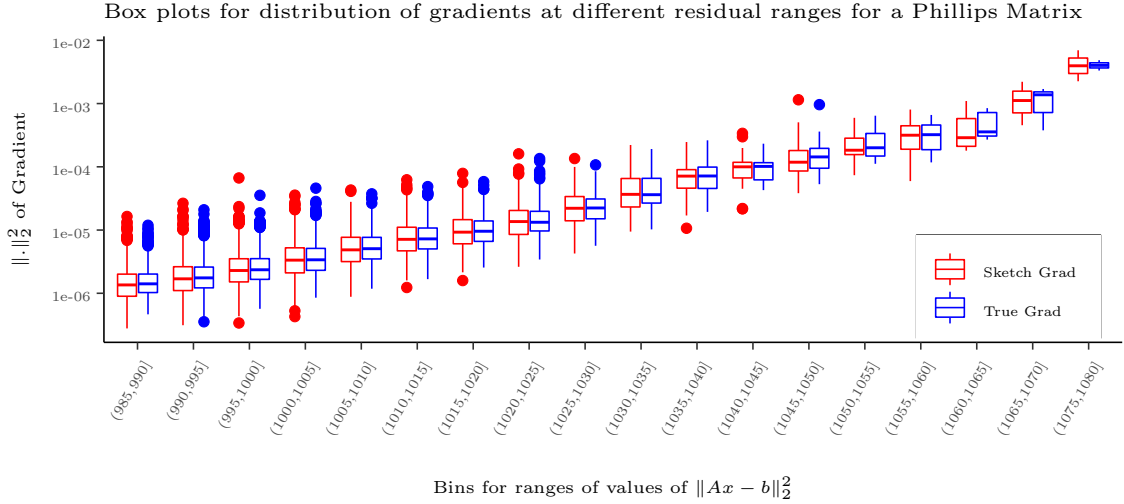}
    \caption{We solve a Phillips linear system, which has a condition number of $\mathcal{O}(10^9)$, from \textit{MatrixDepot} \cite{matrixdepot} using an iterative random sketching method. We compute the norm squared of the true and sketched gradients of the iterates as well as the norm squared of the residual of the iterates. The box plots show the distribution of gradient values for the norms squared of the sketched and true gradients at different intervals of residual norm squared values. For instance, the red box plot and blue box plot over $(985,990]$ represent the distribution of the norms squared of the sketched and true gradients that correspond to residual norm squared values between $985$ and $990$. \label{intro_graph}}
    
\end{figure}

When it comes to solving least squares problems, controlling the solver's accuracy depends on tracking the progress of the iterations and defining clear stopping conditions, which are typically achieved by using the norm of the gradient of the least squares subproblem.\footnote{In the case of this manuscript, the least squares problem we are considering is $\min_{x \in \mathbb{R}^n} \|Ax - b\|_B^2$, and thus the gradient is $g_k = A^{\top}B(Ax_k - b)$.} Unfortunately, the gradient of a large least squares problem is calculated by applying a very large matrix in both its original and transposed orientation to a vector---a procedure that is very costly because of its guaranteed violation of the principal of spatial locality for memory accesses \cite{Lam1991TheCP} (excepting the case in which the matrix is symmetric). This issue is further exacerbated for a randomized solver: the gradient at an iterate of a randomized solver \textit{is never explicitly calculated}, and, even if it were calculated occasionally for monitoring progress, it would be less reliable as we now explain. As the iterates of the randomized solver are random, the gradient evaluated at these iterates inherits this randomness; thus, a wide range of gradient norm values would correspond to the same residual norm squared  in the iterates (see the blue boxes in \cref{intro_graph}), which results in the norm of the gradient being a poor reflection of the residual norm squared. To reiterate, the gradient norm is widely used for tracking and stopping least squares problems, but it is infeasible to calculate for large scale problems, and is unreliable for randomized solvers.

In the class of sketching-based randomized solvers that we consider in this work, the infeasibility of calculating the entire gradient can be addressed by using the sketch of the gradient,\footnote{We can mathematically represent the sketched gradient as $\tilde g_k = S_{k+1}^\top g_k$, where $S_{k+1}$ is a random matrix satisfying properties to be discussed in \cref{sec:algor}.} which \textit{is efficiently and regularly calculated by this class of randomized solvers}.\footnote{While there are cases where this is not true, we generally accept the premise that randomly sketching a matrix can be efficiently calculated. For instance, the Fast Johnson-Lindenstrauss Transform leverages the fast fourier transform to efficiently sketch a matrix \cite{Ailon2009TheFJ}. As another example, a Gaussian sketch can be efficiently applied using emerging photonic hardware, e.g., \url{lighton.ai}.\label{footnote-feasibility-sketching}} However, the sketched gradient norm inherits not only the randomness of the gradient at an iterate, but also the randomness from the sketching procedure. To see this, as shown by the red boxes in \cref{intro_graph}, the sketched gradient norm has an even wider range for the same residual norm squared relative to the gradient norm. Thus, the sketched gradient norm, though feasibly calculated, is even less reliable for tracking and stopping an iterative randomized least squares solver. 

While the sketched gradient norm alone is insufficient to reliably track and stop the underlying randomized least squares solver, if the sketched gradient norm's uncertainty could be quantified, then we could use this uncertainty set to create risk-informed\footnote{By risk-informed, we mean that the user can specify probabilities for which the tracking metrics and stopping conditions can fail.} metrics for tracking and stopping the corresponding underlying algorithm. In this work, we develop a practical, computationally-efficient method for quantifying the uncertainty set of the norm squared of the sketched gradient, and use it to develop risk-informed methods for tracking and stopping the underlying algorithm. In fact, we take this a step further by generalizing our method to a moving average of the sketched gradients, which turns out to be more reliable. We emphasize that our method, which requires only a small additional computational and memory cost over the solver, will accurately reflect the algorithm's progress based on a user-defined threshold for risk, and will stop the algorithm based on a user-defined threshold for risk.%
 We demonstrate the power of our methodology by solving a $0.78$ TB least squares subproblem arising from the incremental 4D-Var algorithm using only $195$ MB of memory, for which LSQR is infeasible (see \cref{subsec:4dvar}). As a result of our methodology, we are enabling the practical integration of an important class of randomized least squares solvers into algorithms that are widely used in science and engineering, which will support solving larger problems in these fields.
%

Note in our previous work \cite{Pritchard2022}, we developed an analogous procedure for consistent linear systems. While at first glance these procedures seem identical owing to our effort to maintain notational consistency, the procedures and their analysis differ in two fundamental ways. First, the procedure and analysis in \cite{Pritchard2022} relies on consistency, which is not available for the least squares problem. Because of consistency, the procedure in \cite{Pritchard2022} can use left-sketching techniques, which are well studied \cite{ACHLIOPTAS2003671,Ailon2009TheFJ,Gower_2015,Takac_Ric,lacotte2021faster,patel2021implicit,pilanciIHS}. Without consistency the procedure in \cite{Pritchard2022} would fail to reflect the progress of the algorithm because left-sketching fails to adequately solve least squares problems \cite{pilanciIHS,article}. Hence, in this work, the procedure uses the less familiar right-sketching approach. 

Second, as the procedure herein uses right-sketching, we must analyze the procedure using arguments about the column space, rather than arguments about the row space as in \cite{Pritchard2022}. Thus, while we follow a similar sequence of steps to, and replicate notation from, \cite{Pritchard2022}, the underlying concepts in the analysis of the two procedures are rather distinct. 

The remainder of this paper is organized as follows.
In \cref{sec:algor}, we specify the problem that we are solving, the algorithm used to solve this problem, our moving average of the norm squared of the gradient estimator, our estimate of its uncertainty set, and our stopping condition. 
In \cref{sec:conver}, we rigorously establish the foundations of our estimators.
In \cref{sec:exp}, we numerically demonstrate the effectiveness of our estimators and compare our algorithm to a state-of-the-art solver.
In \cref{sec:Con}, we conclude.



\section{Notation} \label{sec:Notation}
We use the following the notation in this work. \\ 

\begin{center}
\begin{longtable}{l p{9cm}} \toprule
	\textbf{Symbol} & \textbf{Description} \\ \midrule
    $A$ & A coefficient matrix in $\mathbb{R}^{m \times n}$.\\
    $B$ & A symmetric positive definite matrix in $\mathbb{R}^{m \times m}$.\\ 
    $b$ & A constant vector in $\mathbb{R}^m$. \\
    $S_k$ & A random matrix in $\mathbb{R}^{n \times p}$ that satisfies 
    the Johnson-Lindenstrauss property (See \cref{def:JL}). \\
    $\|\cdot\|_2$ & The standard Euclidean norm.\\
    $x_k$ & The iterate at iteration $k$.\\ 
    $r_k$ & The residual at iteration $k$, i.e., $r_k = Ax_k - b$.\\
    $g_k$ & The gradient of the least squares problem at iteration $k$, i.e., $g_k = A^\top B(Ax_{k} - b).$\\
    $\tilde{g}_k$ & The sketched gradient of the least squares problem at iteration $k$, i.e., $\tilde A_{k+1}^\top B(Ax_{k} - b)$.\\
    $\mathcal{P}$ & The orthogonal projection matrix onto the range of $B^{1/2}A$.\\
    $\rho_k^\lambda$ & The moving average with window width $\lambda$ of the norms squared of the gradients of the least squares problem.\\
    $\iota_k^\lambda$ & The moving average with window width $\lambda$ of the  norms to the fourth power of the gradients of the least squares problem.\\
    $\tilde \rho_k^\lambda$ & The moving average with window width $\lambda$ of the norms squared of the sketched gradients of the least squares problem.\\
    $\tilde \iota_k^\lambda$ & The moving average with window width $\lambda$ of  the norms to the fourth power of the sketched gradients of the least squares problem.\\
    $\mathbf{SE}(\sigma^2,\omega)$ & A sub-Exponential distribution with variance bounded by $\sigma^2$ and scale 
    parameter $\omega$.\\
    $d$ & Reserved for moments of a distribution.\\
    $\eta$ & User defined constriction parameter used in the calculations of interval width and stopping condition.\\
    $Q_k$ & A matrix with orthonormal columns.\\
    $\tau_\ell$ & Stopping times.\\
    $\xi_{I}$ & User specified control on probability of stopping too late.\\
    $\xi_{II}$ & User specified control on probability of stopping too early.\\
    $\upsilon$ & User specified threshold for a small enough $\rho_k^\lambda$ to warrant stopping.\\
    $\mathbb{E}[\cdot]$ & The expectation operator.\\
    $\mathbb{P}(\cdot)$ & The probability measure.\\
    $\mathcal{F}_{k}$ & The $\sigma$-algebra generated by $S_1,\dots, S_k$\\ 
    \bottomrule
\end{longtable}
\end{center}


\section{Problem Formulation \& Algorithm}\label{sec:algor}
We are interested in solving the following minimization problem
\begin{equation}\label{LS}
   \min_{x \in \mathbb{R}^n} \|Ax - b\|_B^2,
\end{equation}
where $A \in \mathbb{R}^{m \times n}$ is a coefficient matrix; $B\in \mathbb{R}^{m \times m}$ is any symmetric positive definite matrix; $b \in \mathbb{R}^{m}$ is a constant vector; and both $m$ and $n$ are large. Note, we allow $m$ and $n$ to be arbitrary, so our methodology applies to overdetermined, underdetermined, and rank-deficient systems. 
Owing to the size of $A$, we can only access $A$ through matrix-vector multiplications; similarly, \textit{though we will not need it in our algorithm}, we can access $A^\top$ through matrix-vector multiplications, though this would be substantially more expensive owing to the needed memory access pattern \cite{Lam1991TheCP}. For all other operations, we make use of efficiently-computed (see \cref{footnote-feasibility-sketching}), sketches of $A$, which we individually denote by (possibly with a subscript)

\begin{equation}
    \tilde A = AS \in \mathbb{R}^{m \times p},
\end{equation}
where $p$ is generally significantly smaller than $n$ (see \cref{rem:size_p}); and $S \in \mathbb{R}^{n \times p}$ is a random matrix that satisfies the Johnson--Lindenstrauss property \cite{JL} defined in the following manner.

\begin{definition}\label{def:JL}
    A matrix $S \in \mathbb{R}^{n \times p}$ satisfies the Johnson-Lindenstrauss property if there exists constants $C,\omega > 0$ such that for all $\delta \geq 0$ and for any $x \in \mathbb{R}^n$,
    \begin{equation}
    \mathbb{P}\left(|\|Sx\|_2^2 - \|x\|_2^2| > \delta \|x\|_2^2\right) < 2e^{-\min\left\{\frac{Cp\delta^2}{2},\frac{\delta}{2\omega} \right\}}.
    \end{equation}
 \end{definition}

\begin{remark}
In \cref{def:JL}, the constants $C$ and $\omega$ are determined by the method used to generate $S$. There are many choices of these methods, such as sparse Rademacher matrices \cite{ACHLIOPTAS2003671}, Fast Johnson Lindenstrauss Transform (FJLT) \cite{Ailon2009TheFJ}, and Gaussian matrices \cite{Dasgupta2003AnEP,indyk,matouvsek2008variants}. 
Estimates for these constants based on numerical experiments are supplied in \cref{table-JL-constant-values}. 
\end{remark}
\begin{table}[hb!]
    \centering
    \caption{Values of $C$ and $\omega$ in \cref{def:JL} for common sampling methods.}
    \label{table-JL-constant-values}
    \begin{tabular}{lccccc}
          \toprule
           &$C$ & $\omega$\\
           \midrule
          Gaussian Matrix \cite{Dasgupta2003AnEP}  & 1.1 & 0.47\\
           Achlioptas \cite{ACHLIOPTAS2003671}  & 1.16 & 0.46\\
           FJLT \cite{Ailon2009TheFJ} & 0.83 & 0.70\\ 
           \bottomrule
       \end{tabular} 
\end{table}

\begin{remark}
By \cref{def:JL}, $\frac{\Vert Sx \Vert_2^2}{\|x\|_2^2}$ is a sub-Exponential (defined in \cref{def:SE}) random variable with parameters $(1/(Cp),\omega)$ \cite{wainwright_2019}.  
\end{remark}

To solve this problem we will employ an important subclass of generalized column-space descent methods (see \cite{patel2022randomized}), which begins with an iterate $x_0 \in \mathbb{R}^n$ and generates a sequence of iterates, $\lbrace x_k : k \in \mathbb{N} \rbrace$, according to the recursive equation
\begin{align}
    &x_k = x_{k-1} - S_k u_k, \\
    &\text{ where } u_k = \argmin_{u \in \mathbb{R}^p} \|\tilde A_k u - (A x_{k-1} - b)\|_B^2,
\end{align}
and $\tilde{A}_k = S_kA$, which can be computed efficiently (see \cref{footnote-feasibility-sketching}). This update is explicitly given by
\begin{equation}\label{tild_genco_Update}
    x_k = x_{k-1} - S_k(\tilde A_k^\top B\tilde A_k)^\dagger \tilde A_k^\top B(Ax_{k-1} - b),
\end{equation}
where $\dagger$ is the Moore-Penrose pseudoinverse; $\lbrace S_k : k \in \mathbb{N} \rbrace$ are independent, identically distributed matrices satisfying \cref{def:JL}. This form is mathematically equivalent to

\begin{equation}\label{genco_Update}
    x_k = x_{k-1} - S_k(S_k^{\top}A^{\top}BAS_k)^\dagger S_k^{\top}A^{\top}B(Ax_{k-1} - b),
\end{equation}
which is a form that will be useful for proving theory relating to the convergence of \cref{tild_genco_Update}, but which \textbf{we do not explicitly use for the algorithm as $A^\top$ is unfavorable to access for large matrices.}

\begin{algorithm}[hp]
    \caption{Tracking and Stopping for Least Squares}\label{algo:Adaptive-window}
    \begin{algorithmic}[1]
    \begin{small}
    \REQUIRE{$A \in \mathbb{R}^{m \times n}$, $b \in \mathbb{R}^m$, $B^{1/2} \in \mathbb{R}^{m \times m}$, $x_0 \in \mathbb{R}^n$,$\lbrace S_k \rbrace$ satisfying \cref{def:JL}.}
    \REQUIRE{Moving average window widths $\lambda_1 \leq \lambda_2 \in \mathbb{N}$.}
    \REQUIRE{$\alpha > 0, \xi_I \in (0,1) , \xi_{II} \in (0,1) , \delta_I \in (0,1)$, $\delta_{II} > 1$, $\eta \geq 1$, $\upsilon > 0$. }
    \STATE $k \leftarrow 0$, $k' \leftarrow \infty$, $\tilde \rho_0^{*} \leftarrow 0$, $\tilde \iota_{0}^{*} \leftarrow 0$, $\lambda \leftarrow 1$, $FLAG \leftarrow $ \FALSE.
    \WHILE{$k == 0$ \OR $\tilde \rho_{k-1}^\lambda \geq \upsilon$ \OR 
        \begin{align} \sqrt{\tilde \iota_{k-1}^\lambda} \geq \min\Bigg\{&\frac{\lambda (1-\delta_{I})^2\upsilon^2 Cp}{(1 + \log(\lambda))2\log(1/\xi_{I})\sqrt{\tilde \iota_{k-1}^\lambda}}, \frac{\lambda \upsilon (1-\delta_{I})}{2\log(1/\xi_{I}) \omega} ,\notag \\ &\frac{\lambda (\delta_{II}-1)^2\upsilon^2 Cp}{(1 + \log(\lambda))2\log(1/\xi_{II}) \sqrt{ \tilde \iota_{k-1}^\lambda}}, \frac{\lambda \upsilon (\delta_{II} - 1)}{2\log(1/\xi_{II}) \omega} \Bigg\} \nonumber \end{align}}\label{stopping-condition}
    \STATE \# Iteration $k$ \#
    \STATE $r_k \leftarrow B^{1/2}(Ax_{k} - b)$
    \STATE  $\tilde A_{k+1} \leftarrow B^{1/2}A S_{k+1}$
    \STATE $\tilde g_{k} \leftarrow \tilde A^{\top}_{k+1}r_k$
    \IF{$k==0$} \label{update_eq_start}
            \STATE $\lambda \leftarrow 1$
    		\STATE $\tilde \rho_{0}, \tilde \iota_{0} \leftarrow \Vert \tilde g_0 \Vert_2^2, \Vert\tilde g_0 \Vert_2^4$
    \ELSE
    		\IF {(\NOT $FLAG$) \AND $\Vert \tilde g_{k} \Vert^2_2 > \Vert \tilde g_{k-1} \Vert^2_2$}
                \STATE $FLAG \leftarrow$ \TRUE
            \ENDIF
    		\IF {(\NOT $FLAG$) \AND $k < \lambda_1$}
                \STATE $\lambda \leftarrow k + 1$ \label{window_change1}
    			\STATE $\tilde \rho_{k}, \tilde \iota_{k} \leftarrow (k \tilde \rho_{k-1} + \Vert \tilde g_{k} \Vert_2^2)/\lambda, (k \tilde \iota_{k-1} + \Vert \tilde g_{k} \Vert_2^4)/\lambda$ \label{cheapupdate1}
    		\ELSIF{(\NOT $FLAG$) \AND $k \geq \lambda_1$}
                \STATE $\lambda \leftarrow \lambda_1$ \label{window_change2}
    			\STATE $\tilde \rho_{k}, \tilde \iota_{k} \leftarrow (\lambda_1 \tilde \rho_{k-1} + \Vert \tilde g_{k} \Vert_2^2 - \Vert \tilde g_{k-\lambda_1}\Vert_2^2)/\lambda, (\lambda_1 \tilde \iota_{k-1} + \Vert \tilde g_{k} \Vert_2^4 - \Vert \tilde g_{k-\lambda_1}\Vert_2^4)/\lambda$\label{cheapupdate2}
    		\ELSIF{$FLAG$ \AND $\lambda < \lambda_2$}
                \STATE $\lambda \leftarrow \lambda+1$ \label{window_change3}
    			\STATE $\tilde \rho_{k}, \tilde \iota_{k} \leftarrow ((\lambda-1) \tilde \rho_{k-1} + \Vert \tilde g_{k} \Vert_2^2)/\lambda, ((\lambda-1) \tilde \iota_{k-1} + \Vert \tilde g_{k} \Vert_2^4)/\lambda$\label{cheapupdate3}
    		\ELSE
                \STATE $\lambda \leftarrow \lambda_2$ \label{window_change4}
    			\STATE $\tilde \rho_{k},\tilde \iota_{k}  \leftarrow (\lambda_2 \tilde \rho_{k-1} + \Vert \tilde g_{k} \Vert_2^2 - \Vert \tilde g_{k-\lambda_2}\Vert_2^2)/\lambda, (\lambda_2 \tilde \iota_{k-1} + \Vert \tilde g_{k} \Vert_2^4 - \Vert \tilde g_{k-\lambda_2}\Vert_2^4)/\lambda$\label{cheapupdate4}
    		\ENDIF
    \ENDIF
    \STATE Update the estimated $(1-\alpha)$-interval by computing: $$\tilde \rho_{k}^\lambda \pm \max\Bigg(\sqrt{\frac{2 \log(2/\alpha)\tilde \iota_{k}^\lambda(1+\log(\lambda))}{Cp\lambda\eta}},
          \frac{2 \log(2/\alpha)\sqrt{\tilde \iota_{k}^\lambda}\omega}{\lambda \eta}\Bigg)$$\label{Credible-Interval}

    \STATE $u_{k+1} \leftarrow \argmin_{u}\|\tilde A_{k+1} u - r_k\|_2^2$ \quad\# See \cite{MillerGentl} and \cite{hayami}
    \STATE $x_{k+1} \leftarrow x_k - S_k u_{k+1}$

    \STATE $k \leftarrow k + 1$
    \ENDWHILE
    \RETURN{$x_{k}$ \AND estimated $(1-\alpha)$-interval}
    \end{small}
    \end{algorithmic}
\end{algorithm}

Under this formulation, \cref{algo:Adaptive-window} presents our methodology for practically tracking and stopping the progress of least squares solvers of the form \cref{genco_Update} for matrices $\lbrace S_k \rbrace$ that satisfy \cref{def:JL}. \cref{algo:Adaptive-window} has several key components that we explain presently.\footnote{In \cref{algo:Adaptive-window}, we use $\tilde A$ to denote $B^{1/2}AS$, possibly with a subscript. This is done to write \cref{algo:Adaptive-window} in terms of $2$-norms.}
\begin{remunerate}
\item \label{estimators} At each iteration, we compute estimators of two key quantities to determine the progress and uncertainty of the algorithm. One quantity we wish to estimate is the moving average of the norms squared of the gradients, $\rho_k^\lambda$, which we define as 
\begin{equation}
    \rho_k^\lambda = \sum_{i = k - \lambda + 1}^{k} \frac{\|g_i\|_2^2}{\lambda},
\end{equation} 
where $g_k = A^{\top}B(Ax_{k} - b)$ is the gradient at iterate $x_k$; and where $\lambda$ is the width of the moving window. When $\lambda=1$, we recover just the gradient at iteration $x_{k-1}$, and, when $\lambda > 1$, we have a moving average of the gradients. As it is infeasible to calculate  $\rho_k^\lambda$, we estimate  $\rho_k^\lambda$ with the norms squared of the sketched gradients \textit{that have already been computed in the updates of our algorithm} (see \cref{tild_genco_Update}),
\begin{equation}
    \tilde \rho_k^\lambda  = \sum_{i=k - \lambda + 1}^k \frac{\|\tilde g_i\|_2^2}{\lambda},
\end{equation}
where $\tilde g_k = \tilde A_{k+1}^{\top}B(Ax_{k} - b)$. When $\lambda=1$, we recover the sketched gradient norm at iterate $x_{k}$, and, when $\lambda > 1$, we have a moving average of the sketched gradient norms, which turns out to be more reliable.

\item We derive a distribution for $\tilde \rho_k^\lambda$ in \cref{subsection:theoretical-ci-sc}. This distribution relies on an unknown quantity that we estimate using 

\begin{equation}\label{tildiota}
    \tilde \iota_k^\lambda  = \sum_{i=k - \lambda + 1}^k \frac{\|\tilde g_i\|_2^4}{\lambda}.
\end{equation} 

\item The matrix $B^{1/2}$ is the square root of the positive definite matrix, $B\in \mathbb{R}^{m \times m}$, used in the general norm. In practice, $B^{1/2}$ can be computed using the Cholesky decomposition, if $B$ is not too dense or large. Fortunately, in many problems that we consider, such as 4D-Var, $B$ has an underlying structure that can be exploited to efficiently compute $B^{1/2}$.

\item The constants $C, \omega, \text{ and } p$ play an important role in the algorithm owing to their relationship with \cref{def:JL}. The parameters $C$ and $\omega$ are constants relating to the size of the tail bound described in \cref{def:JL}, 
which depend on the chosen sketching method, can be found in \cref{table-JL-constant-values}.
The constant $p$ is the embedding dimension of the random matrix $S_k$, and also appears in the tail bound of \cref{def:JL}. A small lower bound on the size of $p$  is necessary for convergence (see \cref{lemm:finite-stopping}, \cref{rem:size_p}).

\item \label{stopping} Line \ref{stopping-condition} contains the conditions for stopping the algorithm. If $\rho_k^\lambda$ could be practically calculated, then the algorithm could be stopped when $\rho_k^\lambda$ falls below a user-specified threshold, $\upsilon$. However, since we must instead use the estimator of $\rho_k^\lambda$, $\tilde \rho_k^\lambda$, stopping when $\tilde \rho_k^\lambda \leq \upsilon$ leads to two possible sources of error.
\begin{remunerate}
\item[5a.] One type of error is associated with stopping the algorithm later than desired. Algorithmically, this scenario arises when $\rho_k^\lambda \leq \upsilon$ while $\tilde \rho_k^\lambda > \upsilon$. To control this error, we need two user-specified quantities. The first quantity specifies how far $\rho_k^\lambda$ is below $\upsilon$. In particular, we let the user choose $\delta_I \in (0,1)$, and we control the probability that $\rho_k^\lambda \leq \delta_{I}\upsilon$ while $\tilde \rho_k^\lambda > \upsilon$. The second quantity is a user specified bound on this probability, $\xi_I$, that indicates the user's level of risk tolerance for possibly stopping too late. 
\item[5b.] The second type of error is associated with stopping too early. Algorithmically, this scenario occurs when $\rho_k^\lambda > \upsilon$, while $\tilde \rho_k^\lambda \leq \upsilon$. Similar to the first scenario, we will let the user choose $\delta_{II} > 1$ to quantify how much larger $\rho_k^\lambda$ is in comparison to $\upsilon$, when $\tilde \rho_k^\lambda < \upsilon$. Then, we control this probability with a user-specified value $\xi_{II}$, which reflects the user's level of risk tolerance for potentially stopping too early. 
\end{remunerate}

\item[6.] The user-specified parameter $\eta$ is an optional parameter to adjust for the conservativeness of the theoretical confidence interval and stopping condition. If the user specifies $\eta = 1$, then there is no adjustment. 
Reasonable, yet still conservative choices for $\eta$ can be found in \cref{table-eta}, which are based on numerical simulations.

\item[7.] Lines \ref{window_change1}, \ref{window_change2}, \ref{window_change3}, and \ref{window_change4} adaptively change the window width of the moving average. This procedure is necessary as there are two distinct phases of convergence in the algorithm. In the first phase, the iterates converge rapidly towards the solution, which necessitates a smaller moving average window width to reduce the impact of earlier iterates. In the second phase, the iterates begin to make less progress and the randomness of the algorithm is more pronounced in their behavior, which necessitates a larger moving average window width to smooth out this randomness. We identify the change point between the two phases to be the iteration where the norm of the sketched gradients are no longer monotonically decreasing, i.e., $\|\tilde g_{k}\|_2^2 > \|\tilde g_{k-1}\|_2^2 $. At this point we slowly increase the width of the window from the narrow window width, $\lambda_1$, by one at each iteration until it reaches that of the wide window width, $\lambda_2$. While we choose the monotonic condition because of its simplicity and effectiveness, other conditions that attempt to estimate the change point between phases could also be used.

\item[8.] Lines \ref{cheapupdate1}, \ref{cheapupdate2}, \ref{cheapupdate3}, and \ref{cheapupdate4} inexpensively update the estimators $\tilde \rho_k^\lambda$ and $\tilde \iota_k^\lambda$, requiring only four floating point operations to calculate. However, this update can suffer from issues of numerical stability, especially for $\tilde \iota_k^\lambda$. If this is a concern, then $\tilde \rho_k^\lambda$ and $\tilde \iota_k^\lambda$ can be computed in $\mathcal{O}(\lambda_2)$ time simply by taking the mean of the nonzero entries in its storage vector, $\rho$ or $\iota$.

\item[9.] Line \ref{Credible-Interval} describes a $1-\alpha$ credible interval designed to contain  $\rho_k^\lambda$ using the estimators $\tilde \rho_k^\lambda$ and $\tilde \iota_k^\lambda$ computed at iteration $k$. As with the stopping condition, this credible interval is derived in \cref{subsection:estimating-CI-SC} from the tail bounding distribution described in \cref{subsection:theoretical-ci-sc}. The parameter $\alpha$ is selected by the user.

\end{remunerate}
\begin{table}[h!]
    \centering
    \caption{Table of Conservative $\eta$ values for three sampling methods}\label{table-eta}
    \begin{tabular}{lccc}\toprule
        \textbf{Method}& Gaussian & FJLT & Achlioptas \\ \midrule 
        $\mathbf{\eta}$ & $3$ & $4$ & $3$ \\        \bottomrule
    \end{tabular}
\end{table}

\section{Validity of the Credible Interval and Stopping Condition}\label{sec:conver}

With an understanding of the parts of \cref{algo:Adaptive-window}, we must now demonstrate the validity of \cref{algo:Adaptive-window}. In particular, we must show that Line \ref{Credible-Interval} is a valid credible interval for $\rho_{k}^\lambda$, and we must show that Line \ref{stopping-condition} controls the probabilities of the two aforementioned errors at $\xi_I$ and $\xi_{II}$. As both the credible interval and stopping condition depend on $\tilde \rho_k^\lambda$ and $\tilde \iota_k^\lambda$, we will need to establish the validity of these two estimators (i.e., their consistency) in order to establish the validity of the credible interval and stopping condition. In turn, as the consistency of $\tilde \rho_k^\lambda$ and $\tilde \iota_k^\lambda$ depends on the convergence of the iterates, $\lbrace x_k \rbrace$, we show the convergence of the iterates in \cref{subsec:allmom} (specifically, see \cref{thm:convergence-of-moments}). Then, we show that $\tilde \rho_k^\lambda$ and $\tilde \iota_k^\lambda$ are consistent estimators for their respective quantities $\rho_k^\lambda$ and $\iota_k^\lambda$\footnote{This quantity has not yet been defined, but will be defined in \cref{subsection:theoretical-ci-sc}.} by deriving a tail bound for both quantities (see \cref{subsection:theoretical-ci-sc,sub-Gaussian:estimator,thm:cons-of-iota}). Now that we have established the validity of $\tilde \rho_k^\lambda$ and $\tilde \iota_k^\lambda$, we derive the credible interval (see \cref{subsection:theoretical-ci-sc,coll:credible-interval}) and stopping condition (see \cref{subsection:theoretical-ci-sc,corr:stopping}). Both the credible interval and stopping condition require a quantity that is impractical to compute, so we establish that using $\tilde \iota_{k}^\lambda$ as a plug-in estimator for the impractical quantity controls the relative error between the theoretical values for the credible interval and stopping condition, and the versions that use the plug-in estimator (see \cref{subsection:estimating-CI-SC,lemm:relative-error-M-iota}).


\subsection{Convergence of the Iterates}\label{subsec:allmom}

To show that the iterates converge to a solution, it is equivalent to show that the gradient of the least squares problem goes to zero. In turn, if $B$ is the identity matrix, it is equivalent to show that the component of the residual of the linear system in the column space of $A$ goes to zero. For general $B$, an analogous equivalence is established in the following lemma.

\begin{lemma}\label{lemm:rangeBA}
    Let $A\in \mathbb{R}^{m\times n}$, $B\in \mathbb{R}^{m\times m}$ be positive definite, and $x \in \mathbb{R}^n$. Let $\mathcal{P}$ be the orthogonal projection onto $\col(B^{1/2}A)$. Then, the gradient of the least squares problem at $x$, $A^{\top}B(Ax - b) = 0$ if and only if $\mathcal{P} B^{1/2} (Ax - b) = 0$.
\end{lemma}
\begin{proof} 
Let $r = Ax - b$. Suppose $A^{\top}Br = 0$,
        \begin{equation}
            0 = A^{\top}Br= A^{\top}B^{1/2} (\mathcal{P}B^{1/2}r + (I - \mathcal{P})B^{1/2}r)= A^{\top}B^{1/2} \mathcal{P}B^{1/2}r, 
        \end{equation}
        Where the last equality comes from $I - \mathcal{P}$ being an orthogonal projector onto the null space of $A^{\top}B^{1/2}$. Since, $\mathcal{P}B^{1/2}r$ is in the range of $B^{1/2}A$ we know that $A^{\top}B^{1/2} \mathcal{P}B^{1/2}r$ will only be zero when $\mathcal{P} B^{1/2}r = 0$. 

    Now suppose $\mathcal{P}B^{1/2}r = 0$. Then,
    \begin{equation}
        A^{\top}Br = A^{\top}B^{1/2} (\mathcal{P}B^{1/2}r + (I - \mathcal{P})B^{1/2}r) = A^{\top}B^{1/2} (I - \mathcal{P})B^{1/2}r=0,
    \end{equation}
    where the last equality follows from $I - \mathcal{P}$ being an orthogonal projector onto the null space of $A^{\top}B^{1/2}$.
\end{proof}

As the preceding lemma establishes, showing $\lbrace \mathcal{P} B^{1/2} r_{k} \rbrace \to 0$ is equivalent to showing that the iterates converge to a solution. Thus, we establish a recursive relationship between $\mathcal{P}B^{1/2} r_{k}$ and $\mathcal{P} B^{1/2} r_{k-1}$. From  \cref{genco_Update},
\begin{equation}\label{res:update}
    r_k = (I - AS_k(S_k^{\top}A^{\top}BAS_k)^\dagger S_k^{\top}A^{\top}B)r_{k-1}.
\end{equation}
Multiplying both sides by $B^{1/2}$,
\begin{equation}\label{res:tranSupdate}
    B^{1/2}r_k = (I - B^{1/2}AS_k(S_k^{\top}A^{\top}BAS_k)^\dagger S_k^{\top}A^{\top}B^{1/2})B^{1/2}r_{k-1}.
\end{equation}
From here, let $\psi_k = B^{1/2}r_k$. Since $\col(B^{1/2}A) \supset \col(B^{1/2} A S_k )$, multiplying both sides by $\mathcal{P}$ produces 
\begin{equation}\label{res:orthUpdate}
    \mathcal{P} \psi_k = (I - B^{1/2}AS_k(S_k^{\top}A^{\top}BAS_k)^\dagger S_k^{\top}A^{\top}B^{1/2}) \mathcal{P} \psi_{k-1}. 
\end{equation}

Finally, since $B^{1/2}AS_k(S_k^{\top}A^{\top}BAS_k)^\dagger S_k^{\top}A^{\top}B^{1/2}$ is an orthogonal projection matrix,  we can define a matrix $Q_k$ to be the matrix with orthonormal columns that span $\col(B^{1/2}A S_k)$. Then we can write \cref{res:orthUpdate} as
\begin{equation}\label{orthonorm-projection}
    \mathcal{P}\psi_k = (I - Q_kQ_k^{\top})\mathcal{P}\psi_{k-1}.  
\end{equation}

With these relationships and notations established, we now turn to establishing convergence.

\paragraph{Geometric Reduction in Residual Components that lie in Column space of $B^{1/2}A$}
Let $\tau_0 = 0$ and $\tau_1$ being the first iteration where 
\begin{equation}\label{cond:tau}
    \col(Q_1) + \col(Q_2) + \dots + \col(Q_{\tau_1}) = \col(B^{1/2}A),
\end{equation} 
is satisfied. Noting that if \cref{cond:tau} is not satisfied then $\tau_1$ is infinite; otherwise, $\tau_1$ is finite and the following lemma holds.

\begin{lemma}\label{lemm:non-increasing}
    Let $\psi_0 \in \mathbb{R}^m$ and let $\left\{\mathcal{P}\psi_k\right\}$ be generated according to \cref{res:orthUpdate} for $\{S_k : k \in \nat\}$, which are independent and identically distributed random matrices satisfying \cref{def:JL}. On the event, $\left\{\tau_1<\infty\right\}$ there exists a $\gamma_1\in\left(0,1\right)$ such that
    \begin{equation}
        \|\mathcal{P}\psi_{\tau_1}\|_2 \leq \gamma_1 \|\mathcal{P}\psi_0\|_2.
    \end{equation}
\end{lemma}
\begin{proof}
    To prove this, it is only necessary to show that $\gamma_{1}$ exists. First, let $q_{k,1},\dots,q_{k,p}$ denote the columns of $Q_k$. Then we can write $\psi_{\tau_1}$ by \cref{orthonorm-projection} as,
    \begin{equation}
        \mathcal{P}\psi_{\tau_1} = \left[\prod_{k=1}^{\tau_1}\left(I - q_{k,j}q_{k,j}^{\top}\right)\right]\mathcal{P}\psi_0.
    \end{equation}
Since $\mathcal{P}\psi_0 \in \col(B^{1/2}A)$, \cite[Theorem 4.1]{patel2021implicit} implies that there $\exists \gamma_1 \in (0,1)$ that is a function of $\{q_{1,1},q_{1,2},\dots,q_{\tau_1,p}\}$ such that $\|\mathcal{P}\psi_{\tau_1}\|_2 \leq \gamma_1 \|\mathcal{P} \psi_{\tau_0}\|_2$. 
\end{proof}

We can easily repeat this argument for more than just $\tau_1$, in fact when $\{\tau_\ell < \infty\}$, define $\tau_{\ell+1}$ to be the first iteration after $\tau_\ell$ where, 
\begin{equation}
    \col(Q_{\tau_\ell +1}) + \col(Q_{\tau_\ell +2}) +\dots+\col(Q_{\tau_{\ell+1}}) = \col(B^{1/2}A),
\end{equation} 
otherwise let $\tau_{\ell+1}$ be infinite. The preceding argument for the existence of $\gamma_1 \in (0,1)$ will then result in the following corollary. 
\begin{corollary}\label{cor:geomRepeat}
    Let $\psi_0 \in \mathbb{R}^m$ and let $\left\{\mathcal{P}\psi_k\right\}$ be generated according to \cref{res:orthUpdate} for $\{S_k : k \in \nat\}$, which are independent and identically distributed random matrices satisfying \cref{def:JL}. On the event, $\cap_{\ell = 1}^L\left\{\tau_\ell<\infty\right\}$ there exist $\gamma_\ell\in\left(0,1\right)$ for $\ell = 1,\dots,L,$ such that
    \begin{equation}\label{cor:eq:geomRepeat}
        \|\mathcal{P}\psi_{\tau_{L}}\|_2 \leq \left(\prod_{\ell = 1}^L \gamma_\ell \right)\|\mathcal{P}\psi_0\|_2.
    \end{equation}
\end{corollary}

\paragraph{Control of Random Rate and Random Iteration} While appearing to indicate the convergence of the $\mathcal{P}\psi_k$, \cref{cor:geomRepeat} does not guarantee that the portion of the $\psi_k$ in the range of $B^{1/2}A$ converges to 0. This lack of guarantee for convergence arises from two possible points of failure, one being the case where $\gamma_\ell \to 1$ as $\ell \to \infty$ and the other being the case where $\tau_\ell$ is infinite. The following result addresses the former issue using the independence of $\{S_k\}$.
\begin{lemma}\label{lemm:iidgamma}
    Let $\lbrace S_k:k\in \nat \rbrace$ be independent and identically distributed random variables. If for any $\ell \in \mathbb{N}$, $\tau_{\ell}$ is finite, then $\{\tau_j - \tau_{j -1}: j \leq \ell\}$ exist and are independent and identically distributed; and $\{\gamma_j : j \leq \ell\}$ are independent and identically distributed. 
\end{lemma}
\begin{proof}
    When $\tau_{\ell}$ is finite, \cite[Theorem 4.1.3]{durrett2013probability} states that $\lbrace Q_{\tau_{\ell} +1}, \ldots, Q_{\tau_{\ell}+k} \rbrace$ given $\tau_{\ell}$ are independent of $\lbrace Q_{1},\ldots,Q_{\tau_{\ell}} \rbrace$ and are identically distributed to $\lbrace Q_1,\ldots,Q_k \rbrace$  for all $k$. Therefore, $\tau_{\ell+1} - \tau_{\ell}$ and $\tau_1$ are independent and identically distributed. It follows that $\gamma_{\ell}$ are independent and identically distributed. 
\end{proof}

So far, we only know that $\tau_0 = 0$ is finite. Hence, we only know that the random variable $\tau_1 - \tau_0$ exists, but we do not know anything about its finiteness. The next result provides the appropriate remedy. 

\begin{lemma}\label{lemm:finite-stopping}
   Let $\left\{S_k:k \in \nat \right\}$ be independent and identically distributed random variables satisfying \cref{def:JL}. If 
    \begin{equation}\label{dim-lower-bound}
        p > \frac{2\log(2)}{C\delta^2}
    \end{equation}
    for some $\delta \in (2\omega \log(2),1)$\footnote{The implicit restriction on $\omega \leq \frac{1}{2\log(2)}$, poses no real concerns in practice \cref{table-JL-constant-values}.}, then $\exists \pi \in (0,1]$ such that for all $\ell \in \nat$ and $k \geq \rank(A)$,
    \begin{equation}\label{stopping-time-prob}
        \pr(\tau_\ell - \tau_{\ell -1} = k) \leq {k-1 \choose \rank(A) - 1}(1-\pi)^{k - \rank(A)}\pi^{\rank(A)}.
    \end{equation}
\end{lemma}
\begin{proof}
    We begin by verifying that for $z \in \col(B^{1/2}A)$ and $z \ne 0$, then $S^{\top}A^{\top}B^{1/2}z \neq 0$ with some nonzero probability. \cref{def:JL} implies that for any $\delta \in (0,1)$,
    \begin{align}
        \pr(\|S^{\top}A^{\top}B^{1/2}z\|_2^2 > 0) &\geq \pr\left(\left|\|S^{\top}A^{\top}B^{1/2}z\|_2^2 - \|A^{\top}B^{1/2}z\|_2^2\right| \leq \delta \|A^{\top}B^{1/2}z\|_2^2 \right)\\
        &\geq 1 - 2 e^{-\min\left\{ \frac{Cp\delta^2}{2},\frac{\delta}{2\omega}\right\}}.
    \end{align}
    When $\delta \in (2\omega \log(2),1)$ is chosen such that \cref{dim-lower-bound} holds, then $1 - 2 e^{-\min\left\{ \frac{Cp\delta^2}{2},\frac{\delta}{2\omega}\right\}} >0$. Moreover, as this bound is independent of $z \in \col(B^{1/2}A)$, we will refer to the lower bound of $ \pr(\|S^{\top}A^{\top}B^{1/2}z\|_2^2 > 0)$ by $\pi \in (0,1]$  for any $z \neq 0$. Thus, owing to the relationship between $Q_k$ and $\col(B^{1/2}AS_k), \pr(\|Q_k^{\top}z\|_2>0) \geq \pi$ for all $z \ne 0$. 
    
    Given that $\{\col(Q_k) : k \in \nat\}$ are independent and identically distributed, we conclude that the probability that $\col(Q_1) + \dots +\col(Q_{k+1})$ increases in dimension from $\col(Q_1) + \dots +\col(Q_{k})$, when $\dim(\col(Q_1) + \dots +\col(Q_{k+1})) < \rank(A)$ is at least $\pi$. This implies that the probability that the dimension increases $\rank(A)$ times in the first $k$ iterations with $k > \rank(A)$ is dominated by a negative binomial distribution, i.e., for $k \geq \rank(A),$
    \begin{equation}
        \pr(\tau_1  = k) \leq {k-1 \choose \rank(A) - 1}(1-\pi)^{k - \rank(A)}\pi^{\rank(A)}.
    \end{equation} 
    As a result, $\tau_1$ is finite with probability one. The result follows by \cref{lemm:iidgamma}.
\end{proof}
\begin{remark}\label{rem:size_p}
    If $\delta = .7$, for the Gaussian, Achlioptas, and FJLT sampling methods one should choose $p \geq 2$ to satisfy the hypothesis of \cref{lemm:finite-stopping}.
\end{remark}

\paragraph{Convergence of the Moments} With the establishment of the previous lemmas we can now conclude the following theorem. 
\begin{theorem}\label{thm:convergence-of-moments}
    Let $x_0 \in \real^n$ and let $\mathcal{P}$ be the orthogonal projection onto $\col(B^{1/2}A)$. Suppose that $\{S_k : k \in \nat\}$ are independent and identically distributed random variables satisfying \cref{def:JL} and \cref{dim-lower-bound} for some $\delta \in (0,1)$. Let $\lbrace x_k : k\in \mathbb{N} \rbrace$ be generated according to \cref{tild_genco_Update}. Define $\psi_k = B^{1/2}(Ax_k-b)$. Then for any $d\in \nat$, $\ex\left[\|\mathcal{P}\psi_{k} \|_{2}^d \right] \to 0$ and $\ex \left[\|\tilde g_k\|_2^d\right] = \ex\left[\|\tilde A_{k+1}^{\top}B(Ax_k-b)\|_2^d \right] \to 0$ 
    as $k \to \infty$. Furthermore, for any particular $\ell$ we have 
    \begin{equation}
        \ex\left[\|\mathcal{P}\psi_{\tau_\ell} \|_2^d \right] \leq \ex[\gamma_1^d]^\ell \|\mathcal{P}\psi_{0}\|_2^d. 
    \end{equation}
\end{theorem}
\begin{proof}
    It is enough to show that $\ex\left[\|\mathcal{P}\psi_{\tau_\ell} \|_2^d \right] \to 0$ as $k \to \infty$. By \cref{lemm:non-increasing}, $\|\mathcal{P}\psi_k\|_2$ is a non-increasing sequence. Thus, we only need to show a subsequence converges to zero. By \cref{cor:geomRepeat} and \cref{lemm:iidgamma} and \cref{lemm:finite-stopping},
   \begin{equation}
    \ex\left[\|\mathcal{P}\psi_{\tau_\ell} \|_2^d \right] \leq \ex[\gamma_1^d]^\ell \|\mathcal{P}\psi_{0}\|_2^d,
   \end{equation} 
   for all $\ell \in \nat$, where $\ex[\gamma_1^d]<1$. Therefore, as $\ell \to \infty,$ the conclusion follows.
\end{proof}

\subsection{Theoretical Values for the Credible Interval and Stopping Condition} \label{subsection:theoretical-ci-sc}
With convergence in all moments established, we now turn to understanding the distributions of $\tilde \rho_k^\lambda$ and $\tilde \iota_k^\lambda$, in order to validate the estimators as well as derive the stopping condition and credible interval. We begin with an examination of the distribution of $\tilde \rho_k^\lambda$. To perform this examination, it is first important to present the definition of a sub-Exponential distribution for it will be used throughout this subsection. 
\begin{definition}\label{def:SE}
    For a random variable $Y$, with $\ex[Y]= \mu$, $Y-\mu$ follows a sub-Exponential, $\text{SE}(\sigma^2,\omega)$, distribution with parameters $\sigma^2$ and $\omega$ if for all $\delta \geq 0$
    \begin{equation}
        \mathbb{P}\left(|Y - \mu| > \delta \right) \leq 2e^{-\min\left\{\delta^2/(2\sigma^2),\delta/(2\omega) \right\}}.
    \end{equation}
    Equivalently, a random variable $Y - \mu$ is sub-Exponential, $\text{SE}(\sigma^2,\omega)$, if 
    \begin{equation}
        \ex [e^{t(Y-\mu)}] \leq e^{\frac{t^2\sigma^2}{2}},
    \end{equation}
    when $|t| < 1/\omega$ \cite{wainwright_2019}.
\end{definition}

With this definition established, we can note intuitively, if the terms of $\tilde \rho_k^\lambda$ were independent, we would trivially have that $\tilde \rho_k^\lambda$ satisfies \cref{def:SE}. Unfortunately, \textit{they are not independent}. Thus, we innovate the following method to derive the distribution of $\tilde \rho_k^\lambda$ to handle the dependencies, which results in only an additional logarithmic term relative to what would have been the case if the terms had been independent.

\begin{theorem}\label{sub-Gaussian:estimator}
    Suppose the setting of \cref{thm:convergence-of-moments}. Define $\siga$ to be 
    the $\sigma$-algebra generated by $S_1,\dots, S_{k-\lambda+1}$, then 
    \begin{equation}
        \tilde \rho_{k}^\lambda  - \rho_{k}^\lambda \Big| \siga \sim \mathbf{SE}\left(\frac{\bound^4 (1 + \log(\lambda))}{ C p \lambda },\frac{\omega \bound^2}{\lambda} \right),
    \end{equation} 
    where $\bound = \Vert A^{\top}B^{1/2}  \Vert_2 \Vert\mathcal{P}B^{1/2} r_{k - \lambda + 1} \Vert_2$ and $r_{k-\lambda+1} = Ax_{k-\lambda} -b$.
\end{theorem}
\begin{proof}
By induction, we prove, for $|t| \leq \lambda/(\omega \bound^2)$, 
\begin{equation}\label{chernoff-bound}
\mathbb{E}\left[ \left. \prod_{i=k-\lambda+1}^k \exp\left\lbrace\frac{t}{\lambda} \left( \Vert \tilde g_i \Vert_2^2 - \Vert g_i \Vert_2^2 \right) \right\rbrace  \right\vert \siga \right] \leq \exp\left( \frac{t^2 \bound^4}{2 C p \lambda } \sum_{j=1}^\lambda \frac{1}{j} \right),
\end{equation}
where $\bound = \Vert A^{\top}B^{1/2}  \Vert_2 \Vert\mathcal{P}B^{1/2} r_{k - \lambda + 1} \Vert_2$ and the bound on $t$ comes from \cref{lemma:abs_diff1}. We can then use a logarithm to bound the summation. As a result, the sub-Exponential distribution of $\tilde \rho_{k}^\lambda  - \rho_{k}^\lambda$ follows by \cref{def:SE}.

The base case of $\lambda = 1$ follows trivially from $\Vert \tilde g_{k-\lambda+1} \Vert_2^2$  being sub-Exponential. Now assume that the result holds for $k-\lambda+ 1$ to $k-1$. Then,
\begin{align}
&\mathbb{E}\left[ \left. \prod_{i=k - \lambda+1}^k \exp\left\lbrace\frac{t\left( \Vert \tilde g_i \Vert_2^2 - \Vert g_i \Vert_2^2 \right)}{\lambda}  \right\rbrace  \right\vert \siga \right] \\
&= \mathbb{E}\left[ \mathbb{E}\left[\left. \left. \prod_{i=k - \lambda+1}^{k} \exp\left\lbrace\frac{t\left( \Vert \tilde g_i \Vert_2^2 - \Vert g_i \Vert_2^2 \right)}{\lambda}  \right\rbrace \right\vert \mathcal{F}_{k} \right] \right\vert \siga \right] \\
&=\mathbb{E}\left[  \mathbb{E} \left[ \left. \exp\left\lbrace\frac{t\left( \Vert \tilde g_k \Vert_2^2 - \Vert g_k \Vert_2^2 \right)}{\lambda}  \right\rbrace \right\vert \mathcal{F}_{k} \right] \right.  \left. \left.  \prod_{i=k - \lambda+1}^{k-1} \exp\left\lbrace\frac{t \left( \Vert \tilde g_i \Vert_2^2 - \Vert g_i \Vert_2^2 \right)}{\lambda} \right\rbrace  \right\vert \siga \right] \\
&\leq \mathbb{E} \left[ \left. \exp\left\lbrace \frac{t^2 \Vert g_k \Vert_2^4}{2 \lambda^2 C p} \right\rbrace \prod_{i=k - \lambda+1}^{k-1} \exp\left\lbrace\frac{t \left( \Vert \tilde g_i \Vert_2^2 - \Vert g_i \Vert_2^2 \right)}{\lambda}  \right\rbrace  \right\vert \siga \right],
\end{align}
where we have made use of $\Vert \tilde g_k \Vert_2^2$ being sub-Exponential in the ultimate line. Now, applying H\"{o}lder's inequality and the induction hypothesis,
\begin{align}
& \mathbb{E} \left[ \left. \exp\left\lbrace \frac{t^2 \Vert g_k \Vert_2^4}{2 \lambda^2 C p} \right\rbrace \prod_{i=k - \lambda+1}^{k-1} \exp\left\lbrace\frac{t\left( \Vert \tilde g_i \Vert_2^2 - \Vert g_i \Vert_2^2 \right)}{\lambda}  \right\rbrace  \right\vert \siga \right] \\
& \leq \mathbb{E}\left[ \left. \exp\left\lbrace \frac{t^2 \Vert g_k \Vert_2^4}{2 \lambda C p} \right\rbrace \right\vert \siga \right]^\frac{1}{\lambda} \mathbb{E} \left[ \left. \prod_{i=k - \lambda+1}^{k-1} \exp\left\lbrace\frac{t\left( \Vert \tilde g_i \Vert_2^2 - \Vert g_i \Vert_2^2 \right)}{\lambda-1}  \right\rbrace  \right\vert \siga \right]^\frac{ \lambda - 1}{\lambda} \\
& \leq \mathbb{E}\left[ \left. \exp\left\lbrace \frac{t^2 \Vert g_k \Vert_2^4}{2 \lambda C p} \right\rbrace \right\vert \siga \right]^\frac{1}{\lambda}\exp\left\lbrace \frac{t^2 \bound^4}{2 C p (\lambda-1) } \sum_{j=1}^{\lambda-1} \frac{1}{j} \right\rbrace^\frac{ \lambda - 1}{\lambda}. \label{eqn-post-holder}
\end{align}

Now, \cref{lemm:rangeBA,cor:geomRepeat,lemm:iidgamma} imply, with probability one,
\begin{equation}\label{gtoM}
\Vert g_k \Vert_2^4 
\leq \Vert A^{\top}B^{1/2} \Vert_2^4 \Vert \mathcal{P}B^{1/2}r_k \Vert_2^4 
\leq \Vert A^{\top}B^{1/2} \Vert_2^4 \Vert \mathcal{P}B^{1/2}r_{k-\lambda+1}\Vert_2^4 
= \bound^4.
\end{equation}
Since $\bound$ is measurable with respect to $\siga$, we apply the inequality of \cref{gtoM} to \cref{eqn-post-holder} to conclude the proof by induction.
\end{proof}

With the establishment of the distribution around the difference between 
$\tilde \rho_k^\lambda$ and $\rho_k^\lambda$, we also obtain the consistency of $\tilde \rho_k^\lambda$ for $\rho_k^\lambda$ from \cref{sub-Gaussian:estimator} by allowing $k \to \infty$, taking the expectation of the sub-Exponential tail bound \cref{def:SE}, and using the dominated convergence theorem to switch the limit and the integral. With this consistency result, {\bf we conclude that $\tilde \rho_k^\lambda$ is a valid estimator for $\rho_k^\lambda$.}

Just as $\tilde \rho_k^\lambda$ is an estimator for $\rho_k^\lambda$, $\tilde \iota_k^\lambda$ is an estimator for the quantity 
\begin{equation}
    \iota_k^\lambda = \sum_{i = k - \lambda +1}^k \frac{\|A^\top (Ax_i - b)\|_2^4}{\lambda},
\end{equation}
which is impractical to compute.
We now turn to showing the validity of $\tilde \iota_k^\lambda$ as an estimator for $\iota_k^\lambda$. To show the validity of $\tilde \iota_k^\lambda$ we transform $\tilde \iota_k^\lambda - \iota_k^\lambda$ into a form where we can make repeated applications of  \cref{chernoff-bound}. After making these applications, we get the consistency result for $\tilde \iota_k^\lambda$ presented in the following theorem.

\begin{theorem}\label{thm:cons-of-iota}
    Under the conditions of \cref{thm:convergence-of-moments}, we have for $\epsilon > 0$
    \begin{equation} \label{eqn-cons-of-iota}
    \begin{aligned}
        &\pr \left(\left|\tilde \iota_k^\lambda - \iota_k^\lambda\right| > \epsilon \big{\vert} \siga \right) \\
        &\leq 2(1+\lambda)\exp\Bigg(-\min\Bigg(\frac{\epsilon^2 Cp \lambda }{2(2 \bound^2 + \sqrt{\lambda \epsilon})^2\bound^4 (1+ \log(\lambda))},\\
        & \quad \quad \quad \quad \quad \quad \quad\quad \quad \quad\quad \quad \frac{\lambda \epsilon}{2 (2 \bound^2 + \sqrt{\lambda \epsilon})\omega \bound^2}\Bigg)\Bigg),
    \end{aligned}  
    \end{equation} 
    where $\bound = \Vert A^{\top}B^{1/2}  \Vert_2 \Vert\mathcal{P}B^{1/2} r_{k - \lambda+1} \Vert_2$ and $r_{k-\lambda+1} = Ax_{k-\lambda+1} -b$. Thus, as $k\to \infty$, $\tilde \iota_k^\lambda$ is a consistent estimator for $\iota_k^\lambda$.
\end{theorem}
\begin{proof}
    Using the definitions of $\iota_k^\lambda$ and $\tilde \iota_k^\lambda$ we have
    \begin{align}
        &\pr\left(|\tilde \iota_k^\lambda - \iota_k^\lambda| > \epsilon\bigg| \siga\right)\\
         &\quad \quad=\pr\left(\left|\sum_{i=k-\lambda + 1}^k\frac{ \|\tilde g_i\|_2^4 - \|g_i\|_2^4}{\lambda} \right|> \epsilon \bigg| \siga\right)  \\
        &\quad \quad \leq \pr\left(\sum_{i=k-\lambda + 1}^k \left|\frac{ \|\tilde g_i\|_2^4 - \|g_i\|_2^4}{\lambda} \right|> \epsilon \bigg| \siga\right) \\
        & \quad \quad  \leq \pr\left(\sum_{i=k-\lambda + 1}^k \left|\frac{ \|\tilde g_i\|_2^2 - \|g_i\|_2^2}{\lambda} \right| \left| \|\tilde g_i\|_2^2 + \|g_i\|_2^2\right|> \epsilon \bigg| \siga\right)\label{eq:whole}.
    \end{align}
    Then, using any constant $G > 2\bound^2$, we partition \cref{eq:whole} into disjoint sets. Thus,
    \begin{small}
    \begin{align}
        &\pr\left(\sum_{i=k-\lambda + 1}^k \left|\frac{ \|\tilde g_i\|_2^2 - \|g_i\|_2^2}{\lambda} \right| \left| \|\tilde g_i\|_2^2 + \|g_i\|_2^2\right|> \epsilon \bigg| \siga\right) \\  
        &=\pr\bigg(\sum_{i=k-\lambda + 1}^k \left|\frac{ \|\tilde g_i\|_2^2 - \|g_i\|_2^2}{\lambda} \right| \left| \|\tilde g_i\|_2^2 + \|g_i\|_2^2\right|> \epsilon , 
        \bigcap_{i=k-\lambda + 1}^k \left\{\left| \|\tilde g_i\|_2^2 + \|g_i\|_2^2\right|\leq G\right\} \bigg| \siga\bigg) \\
         & +\pr\bigg(\sum_{i=k-\lambda + 1}^k \left|\frac{ \|\tilde g_i\|_2^2 - \|g_i\|_2^2}{\lambda} \right| \left| \|\tilde g_i\|_2^2 + \|g_i\|_2^2\right|> \epsilon , \bigcup_{i=k-\lambda + 1}^k \left\{\left| \|\tilde g_i\|_2^2 + \|g_i\|_2^2\right|> G\right\} \bigg| \siga\bigg) \notag\\
         & \label{int-bound}\leq\pr\bigg(\sum_{i=k-\lambda + 1}^k \left|\frac{ \|\tilde g_i\|_2^2 - \|g_i\|_2^2}{\lambda} \right| > \frac{\epsilon}{G} \bigg| \siga \bigg) + \pr\bigg(\bigcup_{i=k-\lambda + 1}^k \left\{\left| \|\tilde g_i\|_2^2 + \|g_i\|_2^2\right|> G\right\} \bigg| \siga\bigg)
    \end{align}
    \end{small}
    From here we will present the bounds for the left and right terms of \cref{int-bound} separately. For the left-hand term of \cref{int-bound} we use a Chernoff bound and \cref{chernoff-bound} resulting in

    \begin{align}
        \label{rchernoff}\pr\bigg(\sum_{i=k-\lambda + 1}^k \left|\frac{ \|\tilde g_i\|_2^2 - \|g_i\|_2^2}{\lambda} \right| > \frac{\epsilon}{G}\bigg| \siga\bigg)\leq & 2\exp \left(\frac{t^2 \bound^4 (1+ \log(\lambda))}{2Cp\lambda} - \frac{\epsilon t}{G}\right),
    \end{align}
    We next wish to minimize this bound. First note that if unconstrained, this minimization would be achieved by setting  $t = \frac{\epsilon Cp\lambda}{G\bound^4(1+\log(\lambda))}$. However, from \cref{def:SE} we know this Chernoff bound only holds when $0 \leq t \leq \frac{\lambda}{\omega\bound^2}$, thus minimizing this bound requires the consideration of two cases. In the first case we consider when $\frac{\epsilon Cp\lambda}{G\bound^4(1+\log(\lambda))} < \frac{\lambda}{\omega \bound^2}$ resulting in the minimum of the Chernoff bound of the left-hand term of \cref{int-bound} being  
    \begin{equation}
        \label{rmint}2\exp\left( - \frac{\epsilon^2 Cp\lambda}{2G^2 \bound^4 (1+\log(\lambda))}\right).
    \end{equation}
    In the second case we consider when $\frac{\epsilon Cp\lambda}{G\bound^4(1+\log(\lambda))}> \frac{\lambda}{\omega\bound^2}$ and in this case we set $t = \frac{\lambda}{\omega\bound^2}$, resulting in the minimum of the Chernoff bound of the left-hand term of \cref{int-bound} being  
    \begin{equation}
        2\exp\left(-\frac{\epsilon\lambda}{2G\bound^2\omega}\right).
    \end{equation}
    Combining these two cases we get that 
    \begin{align}
        \pr\bigg(&\sum_{i=k-\lambda + 1}^k \left|\frac{ \|\tilde g_i\|_2^2 - \|g_i\|_2^2}{\lambda} \right| > \frac{\epsilon}{G}\bigg| \siga\bigg)\\
        &\leq 2\exp\left(-\min\left(\frac{\epsilon^2 Cp \lambda }{2G^2\bound^4 (1+ \log(\lambda))},\frac{\lambda \epsilon}{2 G\omega \bound^2}\right)\right).
    \end{align} 
    
    We next address the right-hand term of \cref{int-bound} for which we have
    \begin{align}
        &\pr\bigg(\bigcup_{i=k-\lambda + 1}^k \left\{\left|{ \|\tilde g_i\|_2^2 + \|g_i\|_2^2}\right|> G\right\} \bigg| \siga\bigg)  \\
         &\quad \quad  =\pr\bigg(\bigcup_{i=k-\lambda + 1}^k \left\{\left|{ \|\tilde g_i\|_2^2 - \|g_i\|_2^2 + 2 \|g_i\|_2^2}\right|> G \right\}\bigg| \siga\bigg)\\
        &\label{bounddef}\quad \quad \leq \pr\bigg(\bigcup_{i=k-\lambda + 1}^k \left\{\left|\frac{ \|\tilde g_i\|_2^2 - \|g_i\|_2^2}{\bound^2}\right| + 2 \bound^2> G\right\} \bigg| \siga\bigg)  \\
        & \quad \quad \leq \sum_{i=k-\lambda + 1}^k\pr\bigg(\left|{ \|\tilde g_i\|_2^2 - \|g_i\|_2^2}\right|> G - 2\bound^2 \bigg| \siga\bigg)\\
        &\label{lchernoff} \quad \quad \leq 2\lambda \exp \left(\frac{t^2 \bound^4}{2Cp} - t\left(G - 2\bound^2\right)\right),
    \end{align}
    where \cref{bounddef} comes from \cref{gtoM},
    \cref{lchernoff} comes from the Chernoff bound and \cref{chernoff-bound}. We next wish to minimize this bound. First note that if unconstrained, this minimization would be achieved by setting  $t = \frac{Cp(G-2\bound^2)}{\bound^4}$. However, from \cref{def:SE} we know this bound only holds when $0 \leq t \leq \frac{1}{\omega\bound^2}$, thus minimizing this Chernoff bounds requires the consideration of two cases. In the first case, $\frac{Cp(G-2\bound^2)}{\bound^4} < \frac{1}{\omega \bound^2}$, which results in the minimum of the Chernoff bound of the right-hand term of \cref{int-bound} being   
    \begin{equation}
        2\lambda \exp\left(-\frac{Cp(G-2\bound^2)^2}{2\bound^4}\right).
    \end{equation}
    In the second case  $\frac{Cp(G-2\bound^2)}{\bound^2} \geq \frac{1}{\omega \bound^2}$ and in this case we set $t = \frac{\lambda}{\omega \bound^2}$ resulting in the minimum of the Chernoff bound of the right-hand term of \cref{int-bound} being  
    \begin{equation}
        2\lambda \exp\left(-\frac{(G-2\bound^2)}{2\omega\bound^2}\right). 
    \end{equation}
    Combing these two cases gives us that 
    \begin{align}
        \pr\bigg(&\bigcup_{i=k-\lambda + 1}^k \left\{\left|{ \|\tilde g_i\|_2^2 + \|g_i\|_2^2}\right|> G\right\} \bigg| \siga\bigg) \\
        &\leq 2\lambda \exp\left(-\min\left(\frac{Cp(G-2\bound^2)^2}{2\bound^4},\frac{(G-2\bound^2)}{2\omega\bound^2}\right)\right).
    \end{align}

By combing the left-hand and right-hand terms of \cref{int-bound} we get 
\begin{align}
    \pr&\left(|\tilde \iota_k^\lambda - \iota_k^\lambda| > \epsilon\bigg| \siga\right)\\
    &\leq 2\exp\left(-\min\left(\frac{\epsilon^2 Cp \lambda }{2G^2\bound^4 (1+ \log(\lambda))},\frac{\lambda \epsilon}{2 G\omega \bound^2}\right)\right)\\ 
    &\quad \quad \quad \quad + 2\lambda \exp\left(-\min\left(\frac{\left(G - 2\bound^2\right)^2Cp}{2\bound^4}, \frac{G - 2\bound^2}{2\bound^2 \omega}\right)\right),\notag 
\end{align}
 This bound can be tightened by minimizing the bound with respect to $G$. To do this minimization we first note that when $G \geq 2 \bound^2 + \sqrt{\lambda \epsilon} > 2\bound^2$ it is the case that
    \begin{equation}
    \begin{aligned}
        \exp\Bigg(-&\min\left(\frac{\epsilon^2 Cp \lambda }{2G^2\bound^4 (1+ \log(\lambda))},\frac{\lambda \epsilon}{2 G\omega \bound^2}\right)\Bigg)\\
        &\geq \exp\left(-\min\left(\frac{\left(G - 2\bound^2\right)^2Cp}{2\bound^4}, \frac{G - 2\bound^2}{2\bound^2 \omega}\right)\right).
    \end{aligned}
    \end{equation}
    We can then upper bound the right side of \cref{eqn-cons-of-iota} in the following manner,
    \begin{align}
        &\inf_{G > 2\bound^2} 2\exp\left(-\min\left(\frac{\epsilon^2 Cp \lambda }{2G^2\bound^4 (1+ \log(\lambda))},\frac{\lambda \epsilon}{2 G\omega \bound^2}\right)\right)\\ & \quad \quad \quad \quad \quad +2\lambda \exp\left(-\min\left(\frac{\left(G - 2\bound^2\right)^2Cp}{2\bound^4}, \frac{G - 2\bound^2}{2\bound^2 \omega}\right)\right) \notag\\
        &\leq \inf_{G > 2 \bound^2 + \sqrt{\lambda \epsilon}} 2(1+\lambda)\exp\left(-\min\left(\frac{\epsilon^2 Cp \lambda }{2G^2\bound^4 (1+ \log(\lambda))},\frac{\lambda \epsilon}{2 G\omega \bound^2}\right)\right)\label{secondary-iota-bound}\\
        &=2(1+\lambda)\exp\Bigg(-\min\Bigg(\frac{\epsilon^2 Cp \lambda }{2(2 \bound^2 + \sqrt{\lambda \epsilon})^2\bound^4 (1+ \log(\lambda))},\\
        & \quad \quad \quad \quad \quad \quad \quad\quad \quad \quad\quad \quad \frac{\lambda \epsilon}{2 (2 \bound^2 + \sqrt{\lambda \epsilon})\omega \bound^2}\Bigg)\Bigg),\notag
    \end{align}
    where the last line comes from recognizing that \cref{secondary-iota-bound} is monotonically increasing when $G>0$. We then conclude consistency by taking the expectation and the limit as $k\to \infty$ of both sides. Then by using the dominated convergence theorem to switch the expectation and the limit we can then use the fact that \cref{thm:convergence-of-moments} implies that $\bound \to 0$ as $k \to \infty$ to get that the bound converges to zero. This implies the desired consistency result.
\end{proof}

With the consistency and distributional results now established, we conclude that our estimators are valid; thus, we are now able to derive the credible interval\footnote{The proof to this corollary can be found in \cref{proof:cred}.} corresponding to Line \ref{Credible-Interval} and the stopping condition\footnote{The proof to this corollary can be found in \cref{stop:proof}.} corresponding to Line \ref{stopping-condition} of \cref{algo:Adaptive-window}.

\begin{corollary}\label{coll:credible-interval}
    Under the conditions of \cref{thm:convergence-of-moments}, a credible interval of level $1 - \alpha$ for $\tilde \rho_k^\lambda$, corresponding to Line \ref{Credible-Interval} in \cref{algo:Adaptive-window},
    is 
    \begin{align} \label{eq:credible-interval-M}
    &\tilde \rho_k^\lambda \pm \max\Bigg(\sqrt{2 \log(2/\alpha)   \frac{\bound^4(1+\log(\lambda))}{Cp\lambda}},
  2 \log(2/\alpha)  \frac{\bound^2\omega}{\lambda}\Bigg).
    \end{align}
\end{corollary}

\begin{corollary}\label{corr:stopping}
	Let $\xi_I, \xi_{II}, \delta_{I} \in (0,1)$, $\delta_{II} > 1$ and $\upsilon > 0$. 
    Under the conditions of \cref{thm:convergence-of-moments}, the following statements are true.  %

    \begin{equation} \label{eq:actual-typeI-control}
        \begin{aligned}
        &\bound^2 \leq \min \left\{ \frac{\lambda (1-\delta_{I})^2\upsilon^2 Cp}{(1 + \log(\lambda))2\log(1/\xi_{I}) \bound^2}, \frac{\lambda \upsilon (1-\delta_{I})}{2\log(1/\xi_{I}) \omega} \right\}\\
        &\Rightarrow
        \mathbb{P}\left[ \tilde \rho_{k+1}^\lambda > \upsilon, \rho_k^\lambda \leq \delta_I\upsilon \bigg{\vert} \siga \right] < \xi_I,
        \end{aligned}
        \end{equation}
        and
    \begin{equation} \label{eq:actual-typeII-control}
        \begin{aligned}
        &\bound^2 \leq \min \left\{ \frac{\lambda (\delta_{II}-1)^2\upsilon^2 Cp}{(1 + \log(\lambda))2\log(1/\xi_{II}) \bound^2}, \frac{\lambda \upsilon (\delta_{II} - 1)}{2\log(1/\xi_{II}) \omega} \right\}\\
        &\Rightarrow
        \mathbb{P}\left[ \tilde \rho_{k+1}^\lambda \leq  \upsilon, \rho_k > \delta_{II} \upsilon \bigg{\vert} \siga \right] < \xi_{II}.
        \end{aligned}
        \end{equation}
\end{corollary}

\subsection{Estimating the Credible Interval and Stopping Condition} \label{subsection:estimating-CI-SC}
\cref{coll:credible-interval,corr:stopping} provide a well-controlled uncertainty set and stopping condition, yet require knowing $\bound$, which is usually not available. As stated before, \cref{coll:credible-interval,corr:stopping} can be operationalized by replacing $\bound^4$ with $\tilde \iota_{k}^\lambda$. Of course, $\bound^4$ and $\tilde \iota_k^\lambda$ must coincide in some sense in order for this estimation to be valid. Indeed, by \cref{thm:convergence-of-moments,thm:cons-of-iota}, both $\bound^4$ and $\tilde \iota_k^\lambda$ converge to zero as $k \to \infty$, which would allow us to estimate $\bound^4$ with $\tilde \iota_k^\lambda$ to generate consistent estimators. However, we could also estimate $\bound^4$ by $0$ to generate consistent estimators, but these would be uninformative during finite time. Therefore, we must establish that estimating $\bound^4$ by $\tilde \iota_k^\lambda$ is also appropriate within some finite time. In the next result, we establish that the relative error between $\bound^4$ and $\tilde \iota_k^\lambda$ is controlled by a constant (in probability).\footnote{The proof of \cref{lemm:relative-error-M-iota} can be found in \cref{rel:err}.} 

\begin{lemma} \label{lemm:relative-error-M-iota}
    Under the conditions of \cref{thm:convergence-of-moments}, for $\epsilon > 0$, $\bound$ as described in \cref{sub-Gaussian:estimator}, $\tilde \iota_k^\lambda$ as defined in \cref{tildiota}, 
    \begin{equation}
    \begin{aligned}
    &\mathbb{P}\left( \left| \frac{\bound - \tilde \iota_k^\lambda}{\bound}\right| > 1 + \epsilon, \bound^4 \neq 0  \bigg| \siga \right)\\ 
    &\leq 2(1+\lambda)\exp\Bigg(-\min\Bigg(\frac{\epsilon^2 Cp \lambda }{2(2 + \sqrt{\lambda \epsilon})^2 (1+ \log(\lambda))},
         \frac{\lambda \epsilon}{2 (2 + \sqrt{\lambda \epsilon})\omega}\Bigg)\Bigg).
    \end{aligned}
    \end{equation}
\end{lemma}

Owing to \cref{lemm:relative-error-M-iota}, the relative error between $\tilde \iota_k^\lambda$ and $\bound$ is reasonably well controlled for practical purposes. As a result,  we can use $\tilde \iota_k^\lambda$ as a plug-in estimator for $\bound$ in the credible interval, \cref{eq:credible-interval-M}, to produce the estimated credible interval suggested in Line \ref{Credible-Interval} of \cref{algo:Adaptive-window}; and we do the same for the stopping condition controls in \cref{eq:actual-typeI-control,eq:actual-typeII-control} to produce the estimated stopping condition in Line \ref{stopping-condition} of \cref{algo:Adaptive-window}.

\section{Experimental results}\label{sec:exp}

Here, we have two goals. First, we demonstrate the correctness of our theory using numerical simulations. Specifically, we verify the consistency of $\tilde \rho_k^\lambda$ and $\tilde \iota_k^\lambda$ (see \cref{subsec:conc}); we verify the coverage probabilities of the credible intervals (see \cref{subsec:cov}); and we verify the effectiveness and error control for the stopping condition (see \cref{subsec:stop}). Second, we compare our method to state-of-the-art methods on an inner loop of incremental 4D-Var at very large scales (see \cref{subsec:4dvar}). A summary of these experiments can be found in \cref{exp_summ_table}.

\begin{table}[H]
    \scriptsize
    \caption[]{Summary of Experiments} \label{exp_summ_table}
    \begin{tabular}{p{2cm}p{4cm}p{4cm}p{3.5cm}}
        \toprule
        Section & Question being addressed & Matrices Used & Dimensions \\
        \midrule
        \cref{subsec:conc} & Are $\tilde\rho_k^\lambda$ and $\iota_k^\lambda$ consistent estimators? & 44 matrices from \textit{MatrixDepot} & 1024 by 512\\
        \midrule
        \cref{subsec:cov}& Are the $(1-\alpha)$ uncertainty sets of $\tilde\rho_k^\lambda$ actually capturing $(1-\alpha)\%$ of $\rho_k^\lambda$? & Wilkinson, Rohess, and Golub matrices from \textit{MatrixDepot} & 512 by 256\\ 
        \midrule
        \cref{subsec:stop}& Are we stopping the algorithm in accordance with user defined risks? &44 matrices from \textit{MatrixDepot} & 1024 by 512 \\
        \midrule
        \cref{subsec:4dvar}& How does this method work at scale? & Subproblem from Incremental 4-D Var for the Shallow Water Equation &  $2N_cN_t$ by $2N_c$, using $N_c \times N_t = \{20, \dots, 1280 \}  \times \{20 \dots 640\}$\footnote{We are using $\times$ here to denote the Cartesian product between sets.}. Additionally, we consider a 5120000 by 20480 system.\\
        \bottomrule
    \end{tabular}
\end{table}

\subsection{Consistency of Esimators}\label{subsec:conc}
\begin{figure}[htb]
    \centering
    \subfigure{\begin{tikzpicture}
    \begin{axis}[ymode = log, 
        every axis plot/.append style={ultra thick},
        major grid style = {lightgray},
        height = .19 \textheight,
        width = .4 \textwidth,
        xlabel = {Iteration},
        ylabel = {Relative Error},
        legend pos = {south west},
        legend columns = 4,
        legend style = {font = \tiny},
        title = {$\frac{|\tilde{\rho}_k^\lambda - \rho_k^\lambda|}{\rho_k^\lambda}$},
        label style={font=\tiny},
        tick label style={font=\tiny},
        ylabel style = {yshift = -.25cm},
        xlabel style = {yshift = .25cm},
        scaled x ticks=false,
        title style={font=\tiny}]
        \addplot[red] table[x = Iteration, y = 5th, col sep=comma, mark = none]{./csvs/convergedrho.csv}; 
        \addplot[blue] table [x = Iteration, y = 50th, col sep=comma, mark = none]{./csvs/convergedrho.csv};
        \addplot[green!60!black] table [x = Iteration, y = 95th, col sep=comma, mark =none]{./csvs/convergedrho.csv};
        \legend{Min, 50th, Max}
    \end{axis}
\end{tikzpicture}
}
    \subfigure{\begin{tikzpicture}
    \begin{axis}[ymode = log, 
        every axis plot/.append style={ultra thick},
        major grid style = {lightgray},
        height = .19 \textheight,
        width = .4 \textwidth,
        xlabel = {Iteration},
        ylabel = {Relative Error},
        legend pos = {south west},
        legend columns = 4,
        legend style = {font = \tiny},
        title = {$\frac{|\tilde{\iota}_k^\lambda - \iota_k^\lambda|}{\iota_k^\lambda}$},
        label style={font=\tiny},
        ylabel style = {yshift = -.25cm},
        xlabel style = {yshift = .25cm},
        scaled x ticks=false,
        tick label style={font=\tiny},
        title style={font=\tiny}]
        \addplot[red] table[x = Iteration, y = 5th, col sep=comma, mark = none]{./csvs/convergediota.csv}; 
        \addplot[blue] table [x = Iteration, y = 50th, col sep=comma, mark = none]{./csvs/convergediota.csv};
        \addplot[green!60!black] table [x = Iteration, y = 95th, col sep=comma, mark =none]{./csvs/convergediota.csv};
        \legend{Min, 50th, Max}
    \end{axis}
\end{tikzpicture}
}
    \centering
    \subfigure{\begin{tikzpicture}
    \begin{axis}[
      axis y line*=left,
      xlabel={Iteration},
      xlabel near ticks,
      ylabel={$|\tilde \rho_k^\lambda - \rho_k^\lambda|$},
      ylabel near ticks,
      height = .19 \textheight,
      width = .4 \textwidth,
      every axis plot/.append style={ultra thick},
      ymode = log,
      ylabel style = {yshift = -.25cm},
      xlabel style = {yshift = .2cm},
      label style={font=\tiny},
      tick label style={font=\tiny}
    ]
      \addplot[orange] table[x = Iteration, y = rho, col sep=comma, mark = none]{./csvs/iotarho.csv};
       \label{plot_1_y1};
    \end{axis}
    \begin{axis}[
      hide x axis,
      axis y line*=right,
      ylabel={$|\tilde \iota_k^\lambda - \iota_k^\lambda|$},
      ylabel near ticks,
      y label/.append style={rotate=180},
      every axis plot/.append style={ultra thick},
      ymode = log,
      height = .19 \textheight,
      width = .4 \textwidth,
      title = {Estimation Error for $\tilde \rho_k^\lambda$ and $\tilde \iota_k^\lambda$ for Hadamard matrix system},
      legend style = {font = \tiny},
      label style={font=\tiny},
      title style={font=\tiny},
      ylabel style = {yshift = .25cm},
      xlabel style = {yshift = .2cm},  
      tick label style={font=\tiny}
    ]
      \addplot[black] table [x = Iteration, y = iota, col sep=comma, mark = none]{./csvs/iotarho.csv};
       \label{plot_1_y2};
        \addlegendimage{/pgfplots/refstyle=plot_1_y1}\addlegendentry{$|\tilde \iota_k^\lambda - \iota_k^\lambda|$};
        \addlegendimage{/pgfplots/refstyle=plot_1_y2}\addlegendentry{$|\tilde \rho_k^\lambda - \rho_k^\lambda|$};
    \end{axis}
   
  \end{tikzpicture}}
    \caption{The top two plots are relative error plots of the min (red), 50th percentile (blue), and max (green) of the relative error between the estimator and actual value. The top left plot features the relative error of $\tilde \rho_k^\lambda$ across 44 least squares problems solved three times, once using each of the Gaussian, Achlioptas, and FJLT sketching methods. The top right plot features the relative error of $\tilde \iota_k^\lambda$ for those same problems. The bottom plot features the absolute error for $\tilde \rho_k^\lambda$ and $\tilde \iota_k^\lambda$ when applied to the Hadamard matrix system from \textit{MatrixDepot}. \label{fig:rho}}
\end{figure}
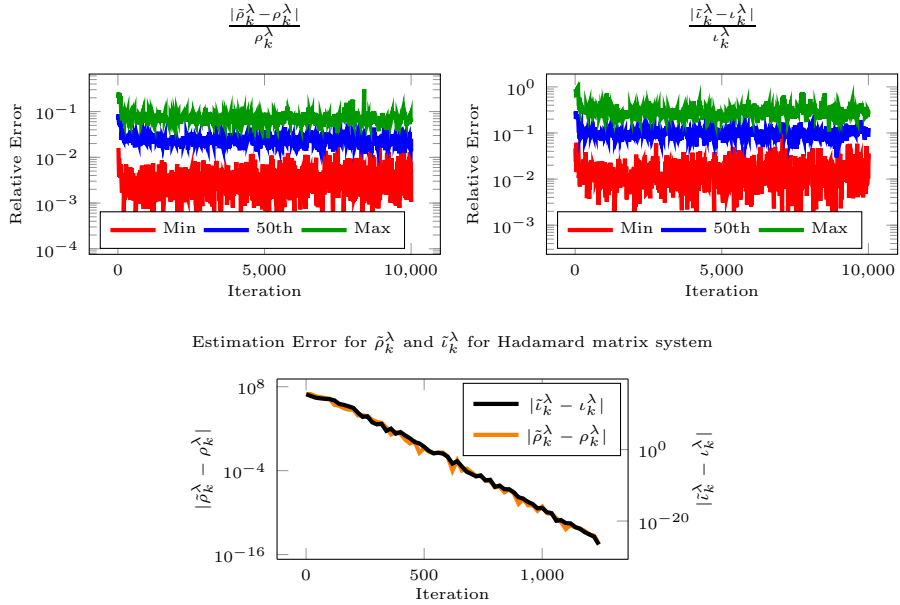

To verify the consistency of our estimators, we solve 44 least squares problems (512 unknowns, 1024 equations) with coefficient matrices generated from \textit{MatrixDepot} \cite{matrixdepot}. Each of these least squares problems is solved three times, once for each of the FJLT, Gaussian, and Achlioptas sketching methods, using an embedding dimension of $p=20$, a narrow moving average window width of $\lambda_1=1$, and a wide moving average window width of $\lambda_2=100$ for 10,000 iterations. At each iteration, for each of the three different sketching methods and 44 matrix systems, the values of $\tilde \rho_k^\lambda$, $\tilde \iota_k^\lambda$, $\rho_k^\lambda$, and $\iota_k^\lambda$ are recorded. Using these values, we compute the relative error for both estimators, $\tilde \rho_k^\lambda$ and $\tilde \iota_k^\lambda$, by taking the absolute value of the difference between the value of the estimator and  the value of the quantity being estimated divided by the value of the quantity being estimated. We then summarize the distribution of these relative errors by computing the min, 50th percentile, and max for both estimator types. In the top left graph of \cref{fig:rho}, we plot these statistics for  the relative error of $\tilde{\rho}_k^\lambda$, $\frac{|\tilde{\rho}_k^\lambda-\rho_k^\lambda|}{\rho_k^\lambda}$; in the top right graph, we do the same for  the relative error of $\tilde \iota_k^\lambda$, $\frac{|\tilde \iota_k^\lambda-\iota_k^\lambda|}{\iota_k^\lambda}$. In the bottom graph we show a specific example of the absolute error $|\tilde \rho_k^\lambda - \rho_k^\lambda|$ (orange line), and the absolute error $|\tilde \iota_k^\lambda - \iota_k^\lambda|$ (black line) for the solver applied to a Hadamard matrix system from \cite{matrixdepot}. More detailed results for the max and min of these relative errors across all sampling types, for each of the 44 systems tested can be found in \cref{sum_tab}.

\begin{table}[ht]
  \scriptsize
  \caption{Max and min relative errors (RE) for $\tilde \iota_k^\lambda$ and $\tilde \rho_k^\lambda$ and condition number for each of the 44 systems across each of the three sampling methods.}\label{sum_tab}
\centering
\begin{tabular}{lrrrrr}
  \hline
Matrix & Condition & Max RE $\tilde \rho_k^\lambda$ & Min RE $\tilde \rho_k^\lambda$ & Max RE $\tilde \iota_k^\lambda$ & Min RE $\tilde \iota_k^\lambda$ \\ 
  \hline
rohess &   1 & 0.63 & 2.6e-06 & 0.93 & 3.6e-05 \\ 
  hadamard &   1 & 0.43 & 1.2e-05 & 0.82 & 1.7e-05 \\ 
  grcar & 3.6 & 0.61 & 2.6e-05 & 1.3 & 3.4e-05 \\ 
  rosser & 3.8 & 0.96 & 1.3e-06 & 2.8 & 7.9e-07 \\ 
  dingdong &   4 & 0.63 & 1.7e-06 & 1.6 & 3.5e-05 \\ 
  parter &   4 & 0.69 & 8.4e-06 & 1.6 & 1.7e-05 \\ 
  randcorr & 4.8 & 0.69 & 7.8e-06 & 0.9 & 2.8e-05 \\ 
  kms &   9 & 0.6 & 3.3e-06 & 1.1 & 4.7e-06 \\ 
  gilbert &  10 & 0.8 & 2e-06 & 2.3 & 3.6e-06 \\ 
  oscillate &  12 & 0.78 & 5.8e-07 & 1.6 & 4.5e-06 \\ 
  smallworld &  34 & 0.59 & 1.3e-06 & 1.3 & 7.8e-06 \\ 
  rando &  76 & 0.71 & 2.1e-06 & 1.2 & 1.1e-05 \\ 
  circul & 5.1e+02 & 0.5 & 4.4e-06 & 1.2 & 5.6e-06 \\ 
  pei & 5.1e+02 & 0.61 & 2.6e-05 & 1.6 & 9.2e-05 \\ 
  hankel & 5.2e+02 & 0.5 & 1.9e-07 & 1.2 & 5.9e-06 \\ 
  wilkinson & 1.2e+03 & 0.47 & 3.8e-08 & 1.2 & 1e-05 \\ 
  randsvd & 4.1e+04 & 0.74 & 6.7e-07 &   2 & 2.2e-05 \\ 
  tridiag & 1.1e+05 & 1.1 & 7e-07 & 3.3 & 3.3e-06 \\ 
  prolate & 1.1e+05 & 0.61 & 1.7e-06 & 1.2 & 5.5e-07 \\ 
  golub & 1.1e+05 & 0.52 & 2e-07 & 1.3 & 1.4e-05 \\ 
  fiedler & 1.8e+05 & 0.76 & 1.4e-06 & 2.1 & 2.8e-07 \\ 
  toeplitz & 1.8e+05 & 0.76 & 4.9e-07 & 2.1 & 1.9e-06 \\ 
  lehmer & 2.8e+05 & 0.7 & 3.6e-07 & 1.9 & 1.4e-05 \\ 
  deriv2 & 3.2e+05 & 0.52 & 7.1e-07 & 1.3 & 4.6e-06 \\ 
  minij & 4.3e+05 & 0.65 & 3.5e-07 & 1.7 & 6e-06 \\ 
  phillips & 1.8e+09 & 0.68 & 9.9e-07 & 1.8 & 1.6e-06 \\ 
  chebspec & 2.2e+14 & 0.68 & 1.6e-06 & 1.8 & 1e-05 \\ 
  ursell & 1.1e+15 & 0.47 & 0.002 & 1.2 & 0.011 \\ 
  chow & 1.2e+16 & 0.6 & 8.6e-07 & 1.5 & 3.5e-06 \\ 
  sampling & 2e+16 & 0.8 & 1.1e-06 & 2.2 & 2.5e-05 \\ 
  moler & 3.2e+17 & 0.57 & 2.4e-07 & 1.5 & 4.5e-06 \\ 
  kahan & 3.8e+17 & 0.42 & 6.6e-07 & 0.92 & 4.3e-06 \\ 
  baart & 4.1e+17 & 0.55 & 5.2e-07 & 1.7 & 3.3e-06 \\ 
  cauchy & 4.6e+18 & 0.8 & 6.2e-06 & 2.2 & 7.8e-06 \\ 
  hilb & 5.8e+18 & 0.88 & 1.6e-06 & 2.5 & 3.5e-05 \\ 
  spikes & 1.4e+19 & 0.79 & 3.6e-06 & 2.2 & 8.8e-06 \\ 
  frank & 1.6e+19 & 0.67 & 2.5e-06 & 1.8 & 1.4e-06 \\ 
  lotkin & 4.2e+19 & 0.7 & 8.1e-06 & 2.2 & 6.3e-06 \\ 
  shaw & 1.7e+20 & 1.1 & 2.1e-06 & 3.4 & 5.1e-06 \\ 
  triw & 2.6e+20 & 0.75 & 3.9e-06 &   1 & 4.6e-06 \\ 
  gravity & 3e+20 & 0.68 & 1.2e-05 & 1.8 & 0.00025 \\ 
  magic & 5.2e+20 & 0.75 & 6.6e-06 & 2.9 & 5.1e-07 \\ 
  foxgood & 1e+21 & 0.54 & 0.074 & 0.79 & 0.15 \\ 
  heat & 8.7e+124 & 0.31 & 4.8e-06 & 0.8 & 5.3e-06 \\ 
   \hline
\end{tabular}
\end{table}

As \cref{sub-Gaussian:estimator,thm:cons-of-iota} show that $\tilde \rho_k^\lambda$ and  $\tilde \iota_k^\lambda$ are consistent estimators, it should be the case that we see constant relative error at all percentiles of the distribution. This is exactly what we obtain when we look at the top two plots in \cref{fig:rho} with all the percentiles corresponding to relative errors that fluctuate around a particular constant. This is confirmed in more detail when looking at \cref{sum_tab} and observing that aside from the Foxgood and Ursell matrices all 44 systems see roughly the same minimum and maximum relative errors for each estimator.  Looking at the bottom graph, which uses the Hadamard matrix system as an illustrative example, we can see that when $\rho_k^\lambda$ converges, the absolute error of $\tilde \rho_k^\lambda$ converges as well. The same is true for when $\iota_k^\lambda$ converges. Overall we can see that our estimators for $\rho_k^\lambda$ and $\iota_k^\lambda$ are quite good performing similarly in terms of relative error of estimators across all systems, and clearly consistent when the value being estimated converges. 

\subsection{Coverage Probability}\label{subsec:cov}
To verify that our credible intervals have the correct coverage probabilities, we perform a two phase experiment where we solve three linear systems (256 unknowns, 512 equations) with coefficient matrices generated from the Golub, Rohess, and Wilkinson matrices found in \textit{MatrixDepot} \cite{matrixdepot}. These matrices are chosen owing to the range of their condition numbers, with the Golub, Rohess, and Wilkison systems having condition numbers of $(81575,1,603)$ respectively, which should help reveal the interplay between the coverage of our intervals and the conditioning of the system. In the first phase of the experiment, we solve each system once for 500 iterations using a Gaussian sketching matrix with an embedding dimension of $p=25$ and a constant moving average window width of $\lambda_1 = \lambda_2 = 15$. At each iteration during this phase, we save the iterate, $\{x_k\}$, $\tilde \rho_k^\lambda$, and the $95\%$ credible interval. Once the first phase is complete we move onto the second phase. The goal of the second phase is to approximate the possible variation in $\rho_{k+15}^\lambda$ given the first phase's iterate, $x_k$ as starting points as dictated by the conditioning in \cref{sub-Gaussian:estimator}. To accomplish this goal for each saved iterate $x_k$ from the first phase, the second phase starts at that $x_k$ and runs \cref{algo:Adaptive-window} for 15 iterations. At each of those 15 iterations, the second phase saves the true norm squared of the gradient, $\|g_k\|_2^2$. At the end of those 15 iterations the second phase uses the fifteen $\|g_k\|_2^2$s to compute $\rho_{k+15}^\lambda$. This process is repeated 1000 independent times for each iterate saved in the first phase. This process results in 1000 observations of $\rho_k^\lambda$ for each iteration greater than 14. Upon the completion of the second phase, we test the coverage of the credible intervals by examining across all iterations how many times the second phase's $\rho_k^\lambda$s exceeded the estimated credible interval from the first phase for its corresponding iteration.

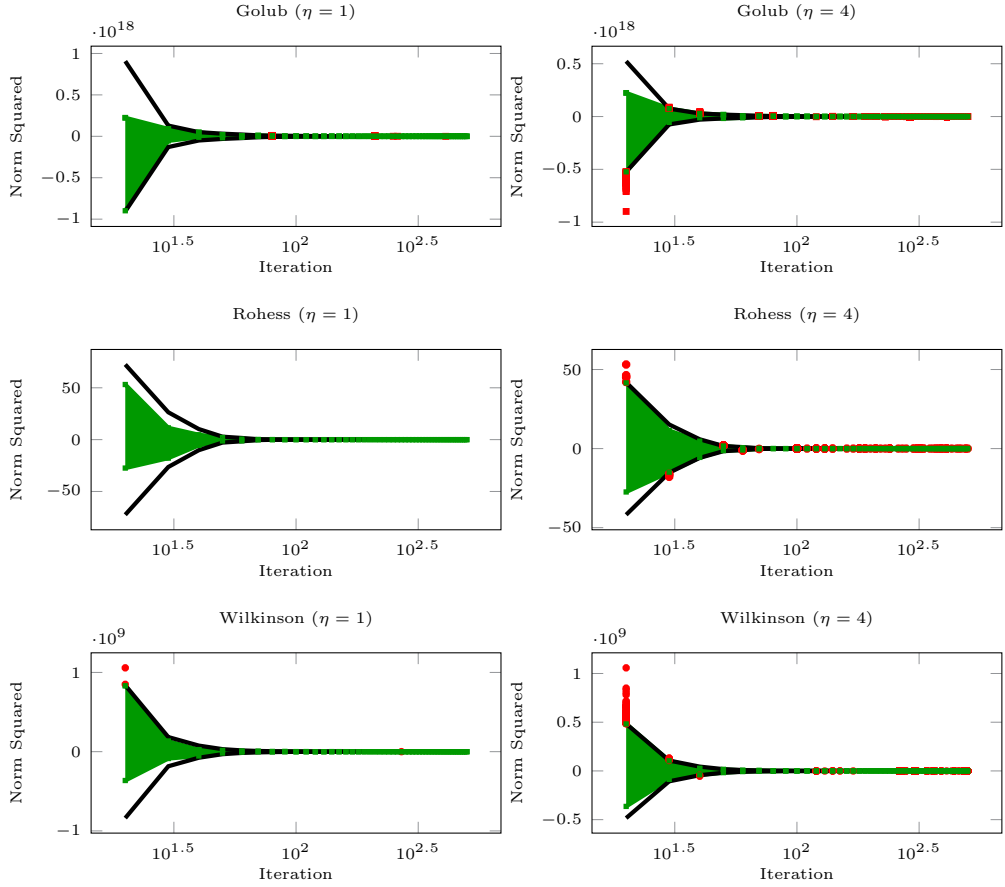
\begin{figure}
    \centering
    \subfigure{\begin{tikzpicture}
    \begin{axis}[
        xmode = log, 
        width = .45 \textwidth,
        height = .19 \textheight,
        every axis plot/.append style={ultra thick},
        xlabel = {Iteration},
        ylabel = {Norm Squared},
        title = {Golub ($\eta = 1$)},
        legend style = {font = \tiny},
      label style={font=\tiny},
      title style={font=\tiny},
      ylabel style = {yshift = -.25cm},
      xlabel style = {yshift = .2cm},  
      tick label style={font=\tiny},
      legend style = {font = \tiny},
      ]
        
        \addplot[green!60!black, only marks, mark = square*, mark size = .3] table [x = it, y = max, col sep=comma]{./csvs/golubrepinside.csv};
        \addplot[red, only marks, mark = square*, mark size = .5] table [x = it, y = Meand, col sep=comma]{./csvs/golubrepoutside.csv};
        \addplot[name path=max,green!60!black, mark = square*, mark size = .2] table [x = it, y = max, col sep=comma]{./csvs/golubrepinside.csv};
        \addplot[name path=min,green!60!black, mark = square*,mark size = .2] table [x = it, y = min, col sep=comma]{./csvs/golubrepinside.csv};
        \addplot[green!60!black] fill between[of=min and max];
        \addplot[black] table[x = it, y = up, col sep=comma, mark = none, tick style = {thin}]{./csvs/golubrepinside.csv}; 
        \addplot[black] table [x = it, y = low, col sep=comma, mark = none]{./csvs/golubrepinside.csv}; 
    \end{axis}
\end{tikzpicture}
\begin{tikzpicture}
    \begin{axis}[
        xmode = log, 
        width = .45 \textwidth,
        height = .19 \textheight,
        every axis plot/.append style={ultra thick},
        xlabel = {Iteration},
        ylabel = {Norm Squared},
        title = {Golub ($\eta = 4$)},
        legend style = {font = \tiny},
      label style={font=\tiny},
      title style={font=\tiny},
      ylabel style = {yshift = -.25cm},
      xlabel style = {yshift = .2cm},  
      tick label style={font=\tiny}]
        
        \addplot[green!60!black, only marks,mark = square*,mark size = .3] table [x = it, y = max, col sep=comma]{./csvs/golubrepinsideeta.csv};
        \addplot[red, only marks,mark = square*, mark size = .5] table [x = it, y = Meand, col sep=comma]{./csvs/golubrepoutsideeta.csv};
        \addplot[name path=max,green!60!black, mark = square*, mark size = .2] table [x = it, y = max, col sep=comma]{./csvs/golubrepinsideeta.csv};
        \addplot[name path=min,green!60!black,  mark = square*, mark size = .2] table [x = it, y = min, col sep=comma]{./csvs/golubrepinsideeta.csv};
        \addplot[green!60!black] fill between[of=min and max]; 
        \addplot[black] table[x = it, y = up, col sep=comma, mark = none]{./csvs/golubrepinsideeta.csv}; 
        \addplot[black] table [x = it, y = low, col sep=comma, mark = none]{./csvs/golubrepinsideeta.csv}; 
        
    \end{axis}
\end{tikzpicture}}
    \subfigure{\begin{tikzpicture}
    \begin{axis}[
        xmode = log,
        width = .45 \textwidth, 
        height = .19 \textheight,
        every axis plot/.append style={ultra thick},
        xlabel = {Iteration},
        ylabel = {Norm Squared},
        title = {Rohess ($\eta = 1$)},
        legend style = {font = \tiny},
      label style={font=\tiny},
      title style={font=\tiny},
      ylabel style = {yshift = -.25cm},
      xlabel style = {yshift = .2cm},  
      tick label style={font=\tiny}]
        
        \addplot[red, only marks, mark size = 1] table [x = it, y = Meand, col sep=comma]{./csvs/rohrepoutside.csv};
        \addplot[name path=max,green!60!black, mark = square*, mark size = .2] table [x = it, y = max, col sep=comma]{./csvs/rohrepinside.csv};
        \addplot[name path=min,green!60!black, mark = square*,mark size = .2] table [x = it, y = min, col sep=comma]{./csvs/rohrepinside.csv};
        \addplot[green!60!black] fill between[of=min and max];
        \addplot[black] table[x = it, y = up, col sep=comma, mark = none]{./csvs/rohrepinside.csv}; 
        \addplot[black] table [x = it, y = low, col sep=comma, mark = none]{./csvs/rohrepinside.csv}; 
        
    \end{axis}
\end{tikzpicture}
\begin{tikzpicture}
    \begin{axis}[
        xmode = log,
        width = .45 \textwidth,
        height = .19 \textheight, 
        every axis plot/.append style={ultra thick},
        xlabel = {Iteration},
        ylabel = {Norm Squared},
        title = {Rohess ($\eta = 4$)},
        legend style = {font = \tiny},
      label style={font=\tiny},
      title style={font=\tiny},
      ylabel style = {yshift = -.25cm},
      xlabel style = {yshift = .2cm},  
      tick label style={font=\tiny}]
        
        \addplot[red, only marks, mark size = .8] table [x = it, y = Meand, col sep=comma]{./csvs/rohessrepoutsideeta.csv};
        \addplot[name path=max,green!60!black, mark = square*, mark size = .2] table [x = it, y = max, col sep=comma]{./csvs/rohessrepinsideeta.csv};
        \addplot[name path=min,green!60!black,  mark = square*, mark size = .2] table [x = it, y = min, col sep=comma]{./csvs/rohessrepinsideeta.csv};
        \addplot[green!60!black] fill between[of=min and max]; 
        \addplot[black] table[x = it, y = up, col sep=comma, mark = none]{./csvs/rohessrepinsideeta.csv}; 
        \addplot[black] table [x = it, y = low, col sep=comma, mark = none]{./csvs/rohessrepinsideeta.csv};
    \end{axis}
\end{tikzpicture}}
    \subfigure{\begin{tikzpicture}
    \begin{axis}[
        xmode = log,
        width = .45 \textwidth,
        height = .19 \textheight,
        every axis plot/.append style={ultra thick},
        xlabel = {Iteration},
        ylabel = {Norm Squared},
        title = {Wilkinson ($\eta = 1$)},
        legend style = {font = \tiny},
      label style={font=\tiny},
      title style={font=\tiny},
      ylabel style = {yshift = -.25cm},
      xlabel style = {yshift = .2cm},  
      tick label style={font=\tiny}]
        
        \addplot[green!60!black, only marks, mark = square*,mark size = .2] table [x = it, y = max, col sep=comma]{./csvs/wilkrepinside.csv};
        \addplot[red, only marks, mark size = .6] table [x = it, y = Meand, col sep=comma]{./csvs/wilkrepoutside.csv};
        \addplot[name path=max,green!60!black, mark = square*, mark size = .2] table [x = it, y = max, col sep=comma]{./csvs/wilkrepinside.csv};
        \addplot[name path=min,green!60!black, mark = square*,mark size = .2] table [x = it, y = min, col sep=comma]{./csvs/wilkrepinside.csv};
        \addplot[green!60!black] fill between[of=min and max];
        \addplot[black] table[x = it, y = up, col sep=comma, mark = none]{./csvs/wilkrepinside.csv}; 
        \addplot[black] table [x = it, y = low, col sep=comma, mark = none]{./csvs/wilkrepinside.csv};
    \end{axis}
\end{tikzpicture}
\begin{tikzpicture}
    \begin{axis}[
        xmode = log,
        width = .45 \textwidth,
        height = .19 \textheight,
        every axis plot/.append style={ultra thick},
        xlabel = {Iteration},
        ylabel = {Norm Squared},
        title = {Wilkinson ($\eta = 4$)},
        legend style = {font = \tiny},
      label style={font=\tiny},
      title style={font=\tiny},
      ylabel style = {yshift = -.25cm},
      xlabel style = {yshift = .2cm},  
      tick label style={font=\tiny}]
        
        \addplot[green!60!black, only marks, mark = square*,mark size = .2] table [x = it, y = max, col sep=comma]{./csvs/wilkrepinsideeta.csv};
        \addplot[red, only marks, mark size = .6] table [x = it, y = Meand, col sep=comma]{./csvs/wilkrepoutsideeta.csv};
        \addplot[name path=max,green!60!black, mark = square*, mark size = .2] table [x = it, y = max, col sep=comma]{./csvs/wilkrepinsideeta.csv};
        \addplot[name path=min,green!60!black,  mark = square*, mark size = .2] table [x = it, y = min, col sep=comma]{./csvs/wilkrepinsideeta.csv};
        \addplot[green!60!black] fill between[of=min and max]; 
        \addplot[black] table[x = it, y = up, col sep=comma, mark = none]{./csvs/wilkrepinsideeta.csv}; 
        \addplot[black] table [x = it, y = low, col sep=comma, mark = none]{./csvs/wilkrepinsideeta.csv};
    \end{axis}
\end{tikzpicture}}
    \caption{Coverage results for credible intervals with $\alpha = 0.05$. The plots on the left display the coverage when the credible interval is calculated with $\eta = 1$, while those on the right are computed with $\eta$ chosen according to \cref{table-eta}. The green points display all the values of $\rho_k^\lambda$ that remain within the interval, while the red points are the values of $\rho_k^\lambda$ that fall outside the interval. The failure rates when $\eta = 1$  for the Golub, Rohess, and  Wilkinson matrices are $(0.00548,0.00000617,0.00121)$ respectively, while when the $\eta$ parameter is set according to \cref{table-eta} these values change to $(0.125,0.0162,0.0428)$.} 
    \label{fig:repsam}
\end{figure}

In \cref{fig:repsam}, we display for each iteration, $k$, the credible interval bound shifted by subtracting the first phase's $\tilde \rho_k^\lambda$, resulting in an interval centered at zero( see black lines). Additionally, for each iteration, we display $\rho_k^\lambda - \tilde \rho_k^\lambda$ for each of the second phase's 1000 different $\rho_k^\lambda$s. If this difference is within the credible interval bound, the observation is colored green otherwise it is colored red. In the left-hand plots of \cref{fig:repsam} we display the results for when the credible interval is computed with $\eta = 1$, while the right-hand plots display the results for a credible interval computed with $\eta$ according to \cref{table-eta}.

From \cref{fig:repsam}, we can observe that with $\eta = 1$ the credible intervals are conservative, with the coverage failure rates of the Golub, Rohess, and  Wilkinson systems being $(0.00548,0.00000617,0.00121)$ respectively all less than the $0.05$ failure rate for which the intervals were designed. With the $\eta$ parameter chosen according to \cref{table-eta}, we observe far less conservative coverage rates across different systems. With the coverage failure rates for the Golub, Rohess, and Wilkinson matrices becoming $(0.0765,0.00565,0.0202)$ respectively. Considering that the condition number for the Golub, Rohess, and Wilkinson matrices are $(81575,1,603)$ these results seem to suggest that the choice in $\eta$ value can be made more or less severe depending on the conditioning of the system, with poorer conditioned systems requiring an $\eta$ value closer to $1$, while better conditioned systems probably require higher $\eta$ values than what is suggested by \cref{table-eta} in order for the intervals to have appropriate coverage rates. Overall, these results demonstrate that, while somewhat conservative, these intervals perform as designed. 

\subsection{Stopping Condition}\label{subsec:stop}
To determine the effectiveness of the stopping condition we again consider 44 least squares problems (512 unknowns, 1024 equations) with coefficient matrices generated from \textit{MatrixDepot} \cite{matrixdepot}. Each of these least squares problems is solved three times for each of the FJLT, Gaussian, and Achlioptas sketching methods with an embedding dimension of $p=20$, a narrow moving average window width of $\lambda_1 = 1$, and a wide moving average window width of $\lambda_2 = 100$ for 10,000 iterations. After solving these systems, we then consider the frequency at which stopping errors of the form of \cref{eq:actual-typeI-control} and \cref{eq:actual-typeII-control} occur when the condition, 
\begin{equation}\label{cond:test}
    \begin{aligned} \sqrt{\tilde \iota_k^\lambda} \leq \min\Bigg\{&\frac{\lambda (1-\delta_{I})^2\upsilon^2 Cp}{(1 + \log(\lambda))2\log(1/\xi_{I})\sqrt{\tilde \iota_k^\lambda}}, \frac{\lambda \upsilon (1-\delta_{I})}{2\log(1/\xi_{I}) \omega} ,\\ &\frac{\lambda (\delta_{II}-1)^2\upsilon^2 Cp}{(1 + \log(\lambda))2\log(1/\xi_{II}) \sqrt{\tilde \iota_k^\lambda}}, \frac{\lambda \upsilon (\delta_{II} - 1)}{2\log(1/\xi_{II}) \omega} \Bigg\} \end{aligned}
\end{equation}
is satisfied. We do this by considering all iterations where \cref{cond:test} is satisfied, then determining the frequency that \cref{eq:actual-typeI-control} ---stopping too late---occurs in these iterations, as well as how often \cref{eq:actual-typeII-control} ---stopping too early---occurs in this set of iterations. The parameters $(\upsilon,\delta_I,\delta_{II}, \xi_I, \xi_{II})$ are set to be $(100,0.9, 1.1, 0.01, 0.01)$.

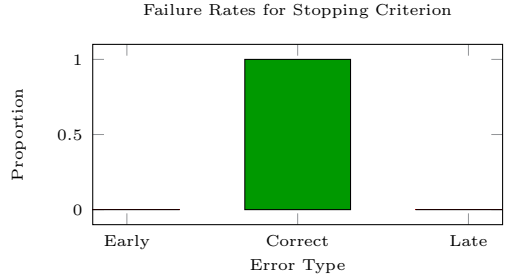
\begin{figure}
    \centering
    \begin{tikzpicture}
    \begin{axis}[
        width = .45 \textwidth,
        height = .19 \textheight,
        xtick={1,2,3},
        xticklabels={Early, Correct, Late},
        xlabel = {Error Type},
        ylabel = {Proportion},
        title = {Failure Rates for Stopping Criterion},
        legend style = {font = \tiny},
      label style={font=\tiny},
      title style={font=\tiny},
      ylabel style = {yshift = -.25cm},
      xlabel style = {yshift = .2cm},  
      tick label style={font=\tiny}, 
        bar width = 40pt]
        \addplot[ybar,  fill = red] coordinates {
            (1, 0)

        };
        \addplot[ybar, fill = green!60!black] coordinates {
            (2, 1)
        };
        \addplot[ybar, fill = red] coordinates {
            (3, 0)
        };
    \end{axis}
\end{tikzpicture}
    \caption{Graph depicting the stopping decision results by error type. The late category describes an error of the form \cref{eq:actual-typeI-control}, while early describes an error of the form  \cref{eq:actual-typeII-control}. These results are displayed with $\eta = 1$; however, they remain unchanged even if $\eta$ is chosen according to \cref{table-eta}.}
    \label{fig:stoppingplot}
\end{figure}

Looking at the \cref{fig:stoppingplot}, we observe that when \cref{cond:test} is satisfied, no error of the form \cref{eq:actual-typeI-control} nor \cref{eq:actual-typeII-control} occurs, and this continues to be the case even with $\eta$ set according to \cref{table-eta}. This low failure rate indicates that overall \cref{cond:test}, stops the algorithm. Thus, if we stop when both $\tilde \rho_k^\lambda \leq \upsilon$ and \cref{cond:test} occur, we will make a stopping decision with a magnitude and error rate acceptable to the user. 

\subsection{4D-Variational Data Assimilation}\label{subsec:4dvar}

To demonstrate the utility of \cref{algo:Adaptive-window} at scale, we consider the Incremental 4D-Variational Data Assimilation problem, 4D-Var \cite{courtier1994strategy}. This problem is solved by iteratively updating an initial estimate by minimizing the distance between noisy observations at different time points and predictions of these observations made by evolving an estimate of the initial state to the same points in time as the observations. To evolve the initial state for our experiment, we use the dynamics defined by the one-dimensional Shallow Water Equations, which are
\begin{align}
\frac{\partial \phi(x,t)}{\partial t} &= - \frac{\partial}{\partial x} \phi(x,t) u(x,t), \text{ and }\label{diffeq:phi1} \\
\frac{\partial u(x,t)}{\partial t} &= - \frac{\partial}{\partial x} \left(\phi(x,t) + \frac{u(x,t)^2}{2}\right), \label{diffeq:u1}
\end{align}
where $x$ is the spatial coordinate; $t$ is the time point; $\phi(x,t)$ is the potential energy; and $u(x,t)$ is the velocity\cite{DimetVar}. 

To solve the 4D-Var problem with these Shallow Water dynamics, rather than directly considering \cref{algo:Adaptive-window}, we consider a modified version of \cref{algo:Adaptive-window}, \cref{alg:4dvar}, specifically tailored to the 4D-Var problem in a way that minimizes memory usage, and compare it to the LSQR solver \cite{lsqr} applied to the same system. For this comparison, we first demonstrate on small problems, those less than 32 GB in size, how \cref{alg:4dvar} produces the same quality of solution as LSQR, has the same runtime scaling as LSQR, and uses substantially less memory than LSQR, when we vary either the number of time points or the number of coordinate points and keep the other at a constant size. We then show the capabilities of \cref{alg:4dvar} exceed those of LSQR, by solving a 4D-Var problem where the system size at 0.78 TB far exceeds 32 GB memory constraint.


To perform both experiments, we generate a set of observations of the potential energy and velocity states for the Shallow Water equations with the desired number of time and coordinate points. This is done using Euler's method with the initial condition of potential energy being set to $\frac{(i-100)^2}{10000}$, where $i$ is the index of the location, and the initial condition on velocity being set to $0.5$ for all coordinates. Each time point is set to be $10^{-11}$ units apart and each coordinate point is 100 units apart to ensure that the system can be stably simulated when the number of coordinates and time points is large. Since in most practical instances, one would only observe either the potential energy or velocity at a particular location, we set all velocity components of the observations to zero. We then add a vector with mean zero, variance one, Gaussian entries to the potential energy states at each time point, which results in our noisy observations.\footnote{Precise formulations of the equations used for Euler's method can be found in \cref{sec:SW}.}

With these observations, we then solve a single inner iteration of the Incremental 4D-Var problem with an initial state estimate of $\frac{(j-100)^4}{10000}$, where $j$ is the entry index of the state vector, once with LSQR and once with \cref{alg:4dvar}. For \cref{alg:4dvar}, we  use the Achlioptas sketching method with an embedding dimension of $p=20$, a narrow moving average window width of $\lambda_1 = 1$, a wide moving average window width of $\lambda_2=100$. In order to account for the floating point errors associated with solving large matrix systems, the threshold for stopping is set to be $ \upsilon = 10^{-9}(N_c(N_t+1))$, where $N_c$ is the number of coordinates and $N_t$ is the number of time points. The other stopping parameters, $(\delta_I,\delta_{II},\xi_I,\xi_{II})$ are set to be $(.9,1.1,.95,.95)$. For both solvers we use a single thread of an Intel Xeon E5-2680 v3 @ 2.50GHz with a memory constraint of 32 GB. We consider systems with the number of time points varying from 20 to 640 by way of doubling, as well as with the number of coordinate points varying from 20 to 1280 by way of doubling. This results in matrix systems that range in size from 250 KB to 31.25 GB.  The LSQR algorithm is stopped once a norm of the gradient of $\sqrt{\upsilon}$ is achieved and \cref{alg:4dvar} is stopped according to Line \ref{stopping-condition} of \cref{algo:Adaptive-window}. Once stopped, we compare the runtime scaling, memory usage, and norm squared of the residual of final solution for both methods in the cases where the number of coordinates changes, but the number of time points stays constant and vice versa.

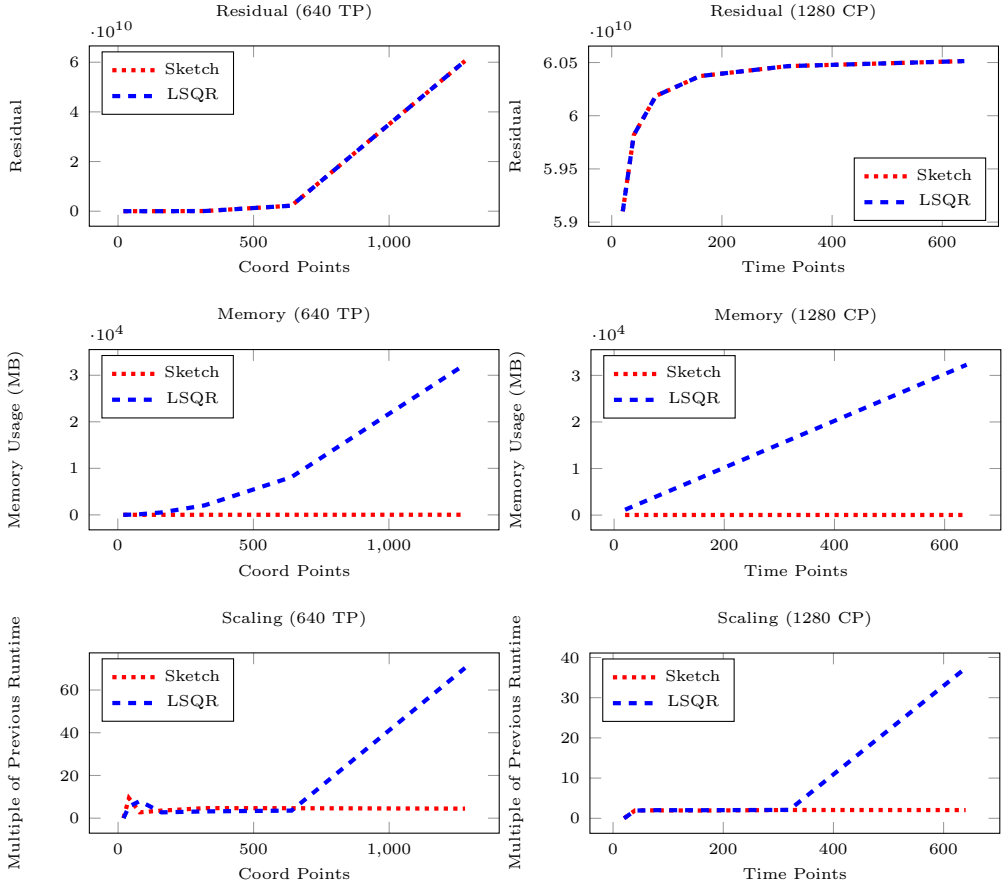
\begin{figure}
    \centering
    \subfigure{\begin{tikzpicture}
    \begin{axis}[
        every axis plot/.append style={ultra thick},
        major grid style = {lightgray},
        height = .19 \textheight,
        width = .45 \textwidth,
        xlabel = {Coord Points},
        ylabel = {Residual},
	    tick label style={font=\small},
        label style={font=\small},
        legend pos = {north west},
        title = {Residual (640 TP)},
        legend style = {font = \tiny},
        label style={font=\tiny},
        title style={font=\tiny},
        ylabel style = {yshift = -.25cm},
        xlabel style = {yshift = .2cm},  
        tick label style={font=\tiny}]
        \addplot[red,dotted] table[x = ncords, y = res_gent, col sep=comma, mark = none]{./csvs/ntime_scale640.csv}; 
        \addplot[blue,dashed] table [x = ncords, y = res_kry, col sep=comma, mark = none]{./csvs/ntime_scale640.csv};
        \legend{Sketch, LSQR}
    \end{axis}
\end{tikzpicture}
\begin{tikzpicture}
    \begin{axis}[ 
        every axis plot/.append style={ultra thick},
        major grid style = {lightgray},
        height = .19 \textheight,
        width = .45 \textwidth,
        xlabel = {Time Points},
        ylabel = {Residual},
        tick label style={font=\small},
        label style={font=\small},
        legend pos = {south east},
        title = {Residual (1280 CP)},
        legend style = {font = \tiny},
        label style={font=\tiny},
        title style={font=\tiny},
        ylabel style = {yshift = -.25cm},
        xlabel style = {yshift = .2cm},  
        tick label style={font=\tiny}]
        \addplot[red,dotted] table[x = ntimes, y = res_gent, col sep=comma, mark = none]{./csvs/coord_scale1280.csv}; 
        \addplot[blue,dashed] table [x = ntimes, y = res_kry, col sep=comma, mark = none]{./csvs/coord_scale1280.csv};
        \legend{Sketch, LSQR}
    \end{axis}
\end{tikzpicture}}
    \subfigure{\begin{tikzpicture}
    \begin{axis}[
        every axis plot/.append style={ultra thick},
        major grid style = {lightgray},
        height = .19 \textheight,
        width = .45 \textwidth,
        xlabel = {Coord Points},
        ylabel = {Memory Usage (MB)},
	    tick label style={font=\small},
        label style={font=\small},
        legend pos = {north west},
        title = {Memory (640 TP)},
        legend style = {font = \tiny},
        label style={font=\tiny},
        title style={font=\tiny},
        ylabel style = {yshift = -.25cm},
        xlabel style = {yshift = .2cm},  
        tick label style={font=\tiny}]
        \addplot[red,dotted] table[x = ncords, y = size_gent, col sep=comma, mark = none]{./csvs/ntime_scale640.csv}; 
        \addplot[blue,dashed] table [x = ncords, y = size_kry, col sep=comma, mark = none]{./csvs/ntime_scale640.csv};
        \legend{Sketch, LSQR}
    \end{axis}
\end{tikzpicture}
\begin{tikzpicture}
    \begin{axis}[ 
        every axis plot/.append style={ultra thick},
        major grid style = {lightgray},
        height = .19 \textheight,
        width = .45 \textwidth,
        xlabel = {Time Points},
        ylabel = {Memory Usage (MB)},
        tick label style={font=\small},
        label style={font=\small},
        legend pos = {north west},
        title = {Memory (1280 CP)},
        legend style = {font = \tiny},
        label style={font=\tiny},
        title style={font=\tiny},
        ylabel style = {yshift = -.25cm},
        xlabel style = {yshift = .2cm},  
        tick label style={font=\tiny}]
        \addplot[red,dotted] table[x = ntimes, y = size_gent, col sep=comma, mark = none]{./csvs/coord_scale1280.csv}; 
        \addplot[blue,dashed] table [x = ntimes, y = size_kry, col sep=comma, mark = none]{./csvs/coord_scale1280.csv}; 
        \legend{Sketch, LSQR }
    \end{axis}
\end{tikzpicture}}
    \subfigure{\begin{tikzpicture}
    \begin{axis}[
        every axis plot/.append style={ultra thick},
        major grid style = {lightgray},
        height = .19 \textheight,
        width = .45 \textwidth,
        xlabel = {Coord Points},
        ylabel = {Multiple of Previous Runtime},
	    tick label style={font=\small},
        label style={font=\small},
        legend pos = {north west},
        title = {Scaling (640 TP)},
        legend style = {font = \tiny},
        label style={font=\tiny},
        title style={font=\tiny},
        ylabel style = {yshift = -.25cm},
        xlabel style = {yshift = .2cm},  
        tick label style={font=\tiny}]
        \addplot[red,dotted] table[x = ncords, y = ntimes_gent, col sep=comma]{./csvs/ntime_scale640.csv}; 
        \addplot[blue, dashed] table [x = ncords, y = ntimes_kry, col sep=comma]{./csvs/ntime_scale640.csv};
        \legend{Sketch, LSQR}
    \end{axis}
\end{tikzpicture}
\begin{tikzpicture}
    \begin{axis}[ 
        every axis plot/.append style={ultra thick},
        major grid style = {lightgray},
        height = .19 \textheight,
        width = .45 \textwidth,
        xlabel = {Time Points},
        ylabel = {Multiple of Previous Runtime},
        tick label style={font=\small},
        label style={font=\small},
        legend pos = {north west},
        title = {Scaling (1280 CP)},
        legend style = {font = \tiny},
        label style={font=\tiny},
        title style={font=\tiny},
        ylabel style = {yshift = -.25cm},
        xlabel style = {yshift = .2cm},  
        tick label style={font=\tiny}]
        \addplot[red,dotted] table[x = ntimes, y = ntimes_gent, col sep=comma, mark = none]{./csvs/coord_scale1280.csv}; 
        \addplot[blue,dashed] table [x = ntimes, y = ntimes_kry, col sep=comma, mark = none]{./csvs/coord_scale1280.csv}; 
        \legend{Sketch, LSQR }
    \end{axis}
\end{tikzpicture}}
    
    \caption{Displays how residual (top), memory (middle), and slowdown (bottom) compare between LSQR and \cref{alg:4dvar}.  The left graphs show scaling when the time points (TP) are set at 640 and the number of coordinates are allowed to vary. The right graphs show scaling when the coordinate points (CP) are set at 1280 and number of time points are allowed to vary. The blue curve shows the results for the LSQR solver while the red line shows the results for our solver implemented with \cref{alg:4dvar}.
    }\label{fig:coord}
\end{figure}

 The results for keeping the number of time points constant at 640 and varying the number of coordinates are displayed on the left of \cref{fig:coord}, and the results for keeping the number of coordinates constant at 1280 and varying the number of time points are displayed on the right of \cref{fig:coord}. In all instances, the minimum residual found by both methods is the same. When considering runtime, the runtime for the LSQR method is faster than \cref{alg:4dvar}, as long as the matrix system size is less than the memory constraint, and if the system size is greater than the memory constraint, the LSQR method fails. Since we care most about how the methods scale with changes in the number of coordinates or number of time points, we present how many times longer the runtime of the solver is at a particular system size, compared to the runtime of the same solver applied to a system with half as many coordinate or time points.

 Looking at both sets of plots, we see that for a fixed time point, the LSQR method and \cref{alg:4dvar} increase at close to the same rate with LSQR taking on average 4.57 times longer to solve a problem with twice as many coordinates and  \cref{alg:4dvar} taking 5 times longer to solve a problem with twice as many coordinates. This trend continues until we reach the system with 1280 coordinate points, at which point the LSQR runtime is 70 times longer than it was at 640 coordinate points, while  \cref{alg:4dvar} only takes 4.46 times longer. Evidence for why \cref{alg:4dvar} does not experience the same scaling issues as LSQR is found in the memory frame of \cref{fig:coord} where we observe the memory usage for \cref{alg:4dvar} remains relatively constant at every value of the number of coordinate points, while the memory usage for LSQR grows quadratically over the same span, reaching a maximum of 31.5 GB of memory used.  A similar story can be observed if we vary the number of time points, with \cref{alg:4dvar} and LSQR algorithm both doubling in runtime for every doubling in the number of time points, until 640 time points are reached, at which point the scaling for LSQR becomes 37 times that of the previous system size, but remains constant for \cref{alg:4dvar}. Overall, we can conclude that to generate the same solution quality, \cref{alg:4dvar} scales as well as LSQR, but with a longer overall runtime. Further, we can say that \cref{alg:4dvar} is significantly more memory efficient than LSQR and is therefore able to avoid the poor scaling effects from memory usage for substantially longer than LSQR.

We finally consider the sketched residual and credible interval for a Shallow Water problem with 250 time points and 10240 spatial coordinates, which equates to a system with a storage requirement of $0.78$ TB. We use Achlioptas sketching with an embedding dimension of $p=20$, a narrow moving average window width of $\lambda_1 = 1$, and a wide moving average window width of $\lambda_2 = 100$. As with the previous problem we solve this system using a single thread of an Intel Xeon E5-2680 v3 @ 2.50GHz with a memory constraint of 32 GB of which \cref{alg:4dvar} uses 194.68 MB.

\begin{figure}
    \centering
        
\begin{tikzpicture}
    \begin{groupplot}[group style ={group size= 2 by 1}, ymode = log,
        width = .4 \textwidth,
        height = .19 \textheight,
        every axis plot/.append style={ultra thick},
        xlabel = {Interation},
        ylabel = {$\tilde \rho_k^\lambda$},
        title = {Credible interval for $\rho_k^\lambda$},
        legend style = {font = \tiny},
        label style={font=\tiny},
        title style={font=\tiny},
        ylabel style = {yshift = -.25cm},
        xlabel style = {yshift = .2cm},  
        tick label style={font=\tiny}]
        \nextgroupplot
        \addplot[black] table[x = Iteration, y = up, col sep=comma, mark = none]{./csvs/big.csv}; 
        \addplot[black] table [x = Iteration, y = low, col sep=comma, mark = none]{./csvs/big.csv};
        \addplot[green!60!black, mark = none] table [x = Iteration, y = rho, col sep=comma]{./csvs/big.csv};
        \draw[black,dashed,thick] (axis cs:50000,10e11) rectangle (axis cs:60000,10e6); 
        \nextgroupplot[xshift = .1 \textwidth]
        \addplot[black, restrict x to domain = 50000:51000] table[x = Iteration, y = up, col sep=comma, mark = none]{./csvs/big.csv}; 
        \addplot[black, restrict x to domain = 50000:51000] table [x = Iteration, y = low, col sep=comma, mark = none]{./csvs/big.csv};
        \addplot[green!60!black, mark = none, restrict x to domain =50000:51000] table [x = Iteration, y = rho, col sep=comma]{./csvs/big.csv};
    \end{groupplot}
    \draw[thick,black,->,shorten >=.08 \textwidth,shorten <=.05\textwidth] 
  		(group c1r1.east) -- (group c2r1.west);
\end{tikzpicture}
    \caption{Displays $\tilde \rho_k^\lambda$ and the credible interval for the single inner iteration solve of the large 4D-Var system which has 10240 spatial coordinates and 250 time points. For better viewing of the interval, the left plot represents all iterations the right plot is simply iterations 50,000 through 51,000.}
    \label{fig:bigexp}
\end{figure}
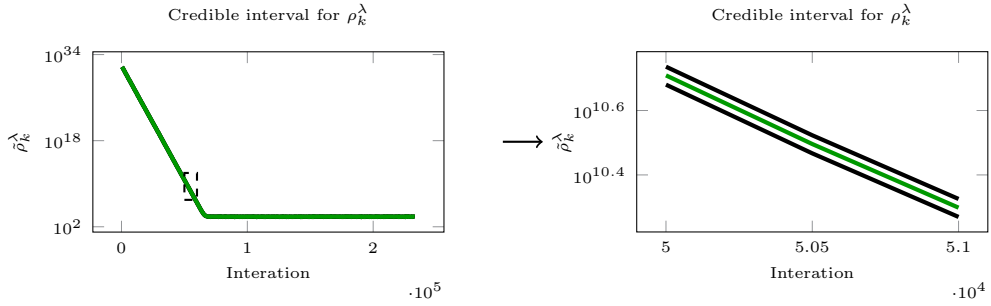

We observe in \cref{fig:bigexp} that most of the progress is made within the first $100,000$ iterations progressing from a $\tilde \rho_k^\lambda$ value of $1.466019\times 10^{32}$ to a value of $6520.793$. The likely cause for this stalled progress is the conditioning of the system, since even at larger sample sizes, $\tilde \rho_k^\lambda$ does not improve beyond $6520.793$. This leads us to claim we have solved the system sufficiently, and have done so under constraints for which LSQR fails to work.


\section{Conclusions}\label{sec:Con}
To efficiently solve the large-scale least squares subproblems that arise in uncertainty quantification, such as 4D-Var, we have proposed an iterative method that leverages random sketching to solve these least squares problems with minimal memory load. The iterative nature of our solution leads to a need to track and stop our method with minimal computational cost, a goal we achieve by utilizing the moving average of the sketched gradients. Through our rigorous proofs, we are then able to verify that not only does our algorithm converge, but also that our estimators are consistent and have a quantifiable uncertainty despite their dependent structure. We perform numerous numerical experiments to verify that this theory holds in practice. In addition to the practical verification of our theory, we make clear the advantages of our method over one like LSQR by comparing both solvers on a 0.78 TB system. Through this comparison, we find that while the LSQR method fails because it reaches the 32 GB memory bound, our method can solve the system utilizing only 195 MB of memory. Our future work will involve improving the practicality of our methodology for solving large-scale scientific problems by examining the effects of the choice of embedding dimension on convergence rate and considering parallelization opportunities to reduce runtime.

\appendix
\section{Commutation of Projection Matrices}
We begin by proving that when one orthogonal projection matrix projections onto a space that is a subset of the space projected onto by a second orthogonal projection matrix, the matrices commute. This allows for us to make the conclusion in \cref{res:orthUpdate}.
\begin{lemma}\label{lemm:commuteproj}
    If $\mathcal{P}_A$ is an orthogonal projection matrix onto the range of $A$ and $\mathcal{P}_{AB}$ is an orthogonal projection matrix onto the range of $AB$, then it is the case
    \begin{equation}
        \mathcal{P}_A \mathcal{P}_{AB} = \mathcal{P}_{AB}\mathcal{P}_A.
    \end{equation}
\end{lemma}
\begin{proof}
We begin by noting that from the properties of orthogonal matrices we have 
    \begin{equation}
        \mathcal{P}_{AB} = \mathcal{P}_A\mathcal{P}_{AB}. \label{project:prop}
    \end{equation}
    \begin{equation}
        \mathcal{P}_{AB} = \mathcal{P}_{AB}^\top = (\mathcal{P}_A  \mathcal{P}_{AB})^\top = \mathcal{P}_{AB}^\top \mathcal{P}_A^\top  =  \mathcal{P}_{AB} \mathcal{P}_{A}, \label{symmetry}
    \end{equation} 
    where the second equality comes from \cref{project:prop} and the second to last equality comes from the symmetry of orthogonal projection matrices. Combining \cref{project:prop,symmetry} we get 
    \begin{equation}
        \mathcal{P}_A\mathcal{P}_{AB} = \mathcal{P}_{AB} = \mathcal{P}_{AB} \mathcal{P}_{A}. 
    \end{equation}

\end{proof}

\section{Scaling of Sub-Exponential distribution}
We present the following useful lemma that is used in the proof of \cref{sub-Gaussian:estimator}. The lemma presents how the distribution of $\|\tilde g_k\|_2^2 - \|g_k\|_2^2 | \mathcal{F}_k$ changes if we scale it by $\|g_k\|_2^2$.
\begin{lemma}\label{lemma:abs_diff1}
    For $x_k$ generated according to \cref{genco_Update} and $S_{k+1}$ satisfying \cref{def:JL}, if we let $\tilde g_k = S_{k+1}^\top A^\top (A x_k - b)$, we let  $\bound = \Vert A^{\top}B^{1/2}  \Vert_2 \Vert\mathcal{P}B^{1/2} r_{k - \lambda + 1} \Vert_2$, and we let $\mathcal{F}_{k}$ be the $\sigma$-algebra generated by $S_1, \dots,S_k$, then
    \begin{equation}
        \|\tilde g_k\|_2^2 - \|g_k\|_2^2 \bigg| \mathcal{F}_{k} \sim \textbf{SE}( \bound^4/(Cp), \omega \bound^2) 
    \end{equation} 
\end{lemma}
\begin{proof}
    We can conclude from \cref{def:SE,def:JL} that if we define $Y = \frac{\|\tilde g_k\|_2^2 - \|g_k\|_2^2}{\|g_k\|_2^2}$, then $Y |\mathcal{F}_{k} \sim \text{SE}(1/(Cp),\omega)$. Thus, we have from \cref{def:SE} that
    \begin{equation}
        \ex\left[\exp\left(tY\right)\right |\mathcal{F}_{k}] \leq \exp\left(\frac{t^2}{2Cp}\right)
    \end{equation}
    when $|t| \leq 1/\omega$. Now, since \cref{lemm:rangeBA,cor:geomRepeat,lemm:iidgamma} imply, with probability one,
    \begin{equation}\label{rtoM}
        \Vert g_k \Vert_2
        \leq \Vert  \Vert A^{\top}B^{1/2}  \Vert_2 \Vert\mathcal{P}B^{1/2} r_{k} \Vert_2
        \leq \Vert  \Vert A^{\top}B^{1/2}  \Vert_2 \Vert\mathcal{P}B^{1/2} r_{k - \lambda + 1} \Vert_2
        = \bound.
    \end{equation}
   If we define a new random variable $Z = \|g_k\|_2^2Y$, we have that
    \begin{align}
        \ex[\exp(tZ)|\mathcal{F}_{k}] &= \ex\left[\exp\left(t\|g_k\|_2^2Y\right)\right|\mathcal{F}_{k}]\\
        &\leq \ex\left[\exp\left(t\bound^2 Y\right)\right|\mathcal{F}_{k}]\\
        &\leq \exp\left(\frac{t^2 \bound^4}{2Cp}\right),
    \end{align}
    where we use the fact that $\bound^4$ is measurable with respect to $\mathcal{F}_{k}$ and can thus be treated as  scaling of $t$. This implies that the above holds for
    \begin{equation}
        \left|t \bound^2\right| \leq 1/\omega, 
    \end{equation}
    which implies $|t| \leq \frac{1}{\omega\bound^2}$. Combining the bound and the constraint on $t$ gives the desired result.
\end{proof}
\section{Details for Credible interval}\label{proof:cred}
We again state \cref{coll:credible-interval} and provide the proof of corollary below.
\begin{corollary}\label{coll:credible-interval1}
    Under the conditions of \cref{thm:convergence-of-moments}, a credible interval of level $1 - \alpha$ for $\tilde \rho_k^\lambda$, corresponding to Line \ref{Credible-Interval} in \cref{algo:Adaptive-window},
    is 
    \begin{align} \label{eq:credible-interval-M1}
    &\tilde \rho_k^\lambda \pm \max\Bigg(\sqrt{2 \log(2/\alpha)   \frac{\bound^4(1+\log(\lambda))}{Cp\lambda}},
  2 \log(2/\alpha)  \frac{\bound^2\omega}{\lambda}\Bigg).
    \end{align}
\end{corollary}
\begin{proof}
    Using the sub-Exponential variance from \cref{sub-Gaussian:estimator}, the tail bound result from Proposition 2.9 of \cite{wainwright_2019},
    \begin{equation}
    \begin{aligned}
        \mathbb{P}&\left(\left| \tilde \rho_{k}^\lambda - \rho_{k}^\lambda \right| > \epsilon | \siga \right)\\ &\leq  2\exp\left(- \min\left(\frac{Cp\lambda\epsilon^2}{2 \bound^4 (1+ \log(\lambda))},\frac{\lambda \epsilon}{2 \omega {\bound^2}}\right)\right).
    \end{aligned}
    \end{equation}
    From this tail probability it should be clear that the high probability region is controlled by the 
    choice of $\epsilon$; thus, if we choose $\epsilon$ such that it lines up with a specific quantile $\alpha$, 
    we will have our desired confidence region.
    \begin{equation}
        \alpha =  2\exp\left(- \min\left(\frac{Cp\lambda\epsilon^2}{2 \bound^4 (1+ \log(\lambda))},\frac{\lambda \epsilon}{2 \omega{\bound^2}}\right)\right). 
    \end{equation}
    Solving for $\epsilon$ in both cases supplies that 
        \begin{equation}
            \epsilon = 
               \max\left(\sqrt{2 \log(2/\alpha)  \frac{\bound^4(1+\log(\lambda))}{Cp\lambda}},
               \frac{2\log(2/\alpha)\omega \bound^2}{\lambda}  \right). 
        \end{equation}
\end{proof}

\section{Details for Stopping Condition}\label{stop:proof}
We again state \cref{corr:stopping} and provide the proof of corollary below.
\begin{corollary}\label{corr:stopping1}
	Let $\xi_I, \xi_{II}, \delta_{I} \in (0,1)$, $\delta_{II} > 1$ and $\upsilon > 0$. 
    Under the conditions of \cref{thm:convergence-of-moments}, the following statements are true.  %
    \begin{equation} \label{eq:actual-typeI-control1}
        \begin{aligned}
        &\bound^2 \leq \min \left\{ \frac{\lambda (1-\delta_{I})^2\upsilon^2 Cp}{(1 + \log(\lambda))2\log(1/\xi_{I}) \bound^2}, \frac{\lambda \upsilon (1-\delta_{I})}{2\log(1/\xi_{I}) \omega} \right\}\\&\Rightarrow
        \mathbb{P}\left[ \tilde \rho_{k+1}^\lambda > \upsilon, \rho_k^\lambda \leq \delta_I\upsilon \bigg{\vert} \siga \right] < \xi_I,
        \end{aligned}
        \end{equation}
        and
    \begin{equation} \label{eq:actual-typeII-control1}
        \begin{aligned}
        &\bound^2 \leq \min \left\{ \frac{\lambda (\delta_{II}-1)^2\upsilon^2 Cp}{(1 + \log(\lambda))2\log(1/\xi_{II}) \bound^2}, \frac{\lambda \upsilon (\delta_{II} - 1)}{2\log(1/\xi_{II}) \omega} \right\}\\
        &\Rightarrow
        \mathbb{P}\left[ \tilde \rho_{k}^\lambda \leq  \upsilon, \rho_k > \delta_{II} \upsilon \bigg{\vert} \siga \right] < \xi_{II}.
        \end{aligned}
        \end{equation}
\end{corollary}
\begin{proof}
    First,
    \begin{align}
    &\mathbb{P}\left( \tilde \rho_k^\lambda > \upsilon, \rho_k^\lambda \leq \delta_I \upsilon \bigg{\vert} \siga \right) \notag \\
    &\leq \mathbb{P}\left( \tilde \rho_k^\lambda - \rho_k^\lambda> \upsilon(1 - \delta_I), \rho_k^\lambda \leq \delta_I \upsilon \bigg{\vert} \siga \right)  \\
    & \leq \mathbb{P}\left(\tilde  \rho_k^\lambda - \rho_k^\lambda> \upsilon(1 - \delta_I) \bigg{\vert} \siga \right)
    \end{align}
        Using the sub-Exponential variance from \cref{sub-Gaussian:estimator} and the tail bound result from Proposition 2.9 of \cite{wainwright_2019},
        \begin{align}
            &\mathbb{P}\left(\tilde \rho_{k}^\lambda - \rho_k^\lambda > \upsilon (1-\delta_I) \bigg{\vert} \siga\right)  \notag \\
            & \quad \quad \leq  \exp\left(- \min\left(\frac{\lambda \upsilon^2 (1-\delta_I)^2Cp}{2 \bound^4 (1+ \log(\lambda))},\frac{\lambda \upsilon (1 - \delta_I)}{2\omega \bound^2}\right)\right),
        \end{align} 
         Thus, when $\bound^2$ satisfies \cref{eq:actual-typeI-control}, the right-hand term of the preceding inequality is bounded by $\xi_I$.
    We can repeat this argument to show that \cref{eq:actual-typeII-control} is true.
    \end{proof}

\section{Relative error bound for $\iota_k^\lambda$}
Here we define the relative error bound to be used in the proof of \cref{lemm:relative-error-M-iota}, which is simply a relative error version of \cref{thm:cons-of-iota}. We present the exact derivation of this bound in the following lemma.

\begin{lemma}\label{lem:rel-of-iota}
    Under the conditions of \cref{thm:convergence-of-moments}, we have for $\epsilon >0$
    \begin{equation}
    \begin{aligned}
        &\pr \left(\frac{\left|\tilde \iota_k^\lambda - \iota_k^\lambda\right|}{\bound^4} > \epsilon \bigg| \siga \right)\\ 
         &\leq  2(1+\lambda)\exp\left(-\min\left( \frac{\epsilon^2 Cp\lambda}{2(2 + \sqrt{\epsilon \lambda})^2 (1+\log(\lambda))},\frac{\epsilon\lambda}{2(2 + \sqrt{\epsilon \lambda})\omega}\right)\right)
    \end{aligned}       
    \end{equation} 
\end{lemma}

\begin{proof}
    Using the definitions of $\iota_k^\lambda$ and $\tilde \iota_k^\lambda$ we have
    \begin{align}
        &\pr\left(\left|\frac{\tilde \iota_k^\lambda - \iota_k^\lambda}{\bound^4}\right| > \epsilon\bigg| \siga\right)\\
         &\quad \quad=\pr\left(\left|\sum_{i=k-\lambda + 1}^k\frac{ \|\tilde g_i\|_2^4 - \|g_i\|_2^4}{\lambda\bound^4} \right|> \epsilon \bigg| \siga\right)  \\
        &\quad \quad \leq \pr\left(\sum_{i=k-\lambda + 1}^k \left|\frac{ \|\tilde g_i\|_2^4 - \|g_i\|_2^4}{\lambda\bound^4} \right|> \epsilon \bigg| \siga\right) \\
        & \quad \quad  \leq \pr\left(\sum_{i=k-\lambda + 1}^k \left|\frac{ \|\tilde g_i\|_2^2 - \|g_i\|_2^2}{\lambda\bound^2} \right| \left|\frac{ \|\tilde g_i\|_2^2 + \|g_i\|_2^2}{\bound^2}\right|> \epsilon \bigg| \siga\right)\label{eq:whole1}.
    \end{align}
    Then by defining a variable, $G > 2$, to partition \cref{eq:whole1} into disjoint sets and using the definition of measure,
    \begin{align}
        &\pr\left(\sum_{i=k-\lambda + 1}^k \left|\frac{ \|\tilde g_i\|_2^2 - \|g_i\|_2^2}{\lambda\bound^2} \right| \left|\frac{ \|\tilde g_i\|_2^2 + \|g_i\|_2^2}{\bound^2}\right|> \epsilon \bigg| \siga\right) \\  
        &\quad \quad =\pr\bigg(\sum_{i=k-\lambda + 1}^k \left|\frac{ \|\tilde g_i\|_2^2 - \|g_i\|_2^2}{\lambda\bound^2} \right| \left|\frac{ \|\tilde g_i\|_2^2 + \|g_i\|_2^2}{\bound^2}\right|> \epsilon , \\ 
         &\quad \quad \quad \quad\bigcap_{i=k-\lambda + 1}^k \left\{\left|\frac{ \|\tilde g_i\|_2^2 + \|g_i\|_2^2}{\bound^2}\right|\leq G\right\} \bigg| \siga\bigg)  \notag\\
         & \quad \quad+\pr\bigg(\sum_{i=k-\lambda + 1}^k \left|\frac{ \|\tilde g_i\|_2^2 - \|g_i\|_2^2}{\lambda\bound^2} \right| \left|\frac{ \|\tilde g_i\|_2^2 + \|g_i\|_2^2}{\bound^2}\right|> \epsilon , \notag\\ 
         &\quad \quad \quad \quad\bigcup_{i=k-\lambda + 1}^k \left\{\left|\frac{ \|\tilde g_i\|_2^2 + \|g_i\|_2^2}{\bound^2}\right|> G\right\} \bigg| \siga\bigg)\notag\\
         & \label{int-bound1} \leq\pr\bigg(\sum_{i=k-\lambda + 1}^k \left|\frac{ \|\tilde g_i\|_2^2 - \|g_i\|_2^2}{\lambda\bound^2} \right| > \frac{\epsilon}{G} \bigg| \siga \bigg)\\ &\quad \quad \quad + \pr\bigg(\bigcup_{i=k-\lambda + 1}^k \left\{\left|\frac{ \|\tilde g_i\|_2^2 + \|g_i\|_2^2}{\bound^2}\right|> G\right\} \bigg| \siga\bigg)\notag.
    \end{align}
    From here we will present the bounds for the left and right terms of \cref{int-bound1} separately. For the left term of \cref{int-bound1} we use \cref{chernoff-bound} and have

    \begin{align}
        \label{rchernoff1}\pr\bigg(\sum_{i=k-\lambda + 1}^k \left|\frac{ \|\tilde g_i\|_2^2 - \|g_i\|_2^2}{\lambda \bound^2} \right| > \frac{\epsilon}{G}\bigg| \siga\bigg)\leq & 2\exp \left(\frac{t^2  (1+ \log(\lambda))}{2Cp\lambda} - \frac{\epsilon t}{G}\right),
    \end{align}
    This bound only holds when $0 \leq t \leq \frac{\lambda}{\omega}$. It is first important to note in the unconstrained case the global minimizer of this function occurs at $t = \frac{\epsilon Cp\lambda}{G(1+\log(\lambda))}$. However, since we have a constraint there are two cases we must consider. For the first case we have $\frac{\epsilon Cp\lambda}{G(1+\log(\lambda))} < \frac{\lambda}{\omega}$, and we get the Chernoff bound to be
    \begin{equation}
        \label{rmint1}2\exp\left( - \frac{\epsilon^2 Cp\lambda}{2G^2 (1+\log(\lambda))}\right).
    \end{equation}
    In the second case, $\frac{\epsilon Cp\lambda}{G(1+\log(\lambda))}> \frac{\lambda}{\omega}$ and in this case we minimize the function by setting $t = \frac{\lambda}{\omega}$, which causes the Chernoff bound to be
    \begin{equation}
        2\exp\left(-\frac{\epsilon\lambda}{2G\omega}\right).
    \end{equation}
    Combining these two cases we get that 
    \begin{align}
        \pr\bigg(&\sum_{i=k-\lambda + 1}^k \left|\frac{ \|\tilde g_i\|_2^2 - \|g_i\|_2^2}{\lambda \bound^2} \right| > \frac{\epsilon}{G}\bigg| \siga\bigg)\\
        &\leq2\exp\left(-\min\left( \frac{\epsilon^2 Cp\lambda}{2G^2 (1+\log(\lambda))},\frac{\epsilon\lambda}{2G\omega}\right)\right) .
    \end{align} 
    
    We next address the right side of \cref{int-bound1} for which we have
    \begin{align}
        &\pr\bigg(\bigcup_{i=k-\lambda + 1}^k \left\{\left|\frac{ \|\tilde g_i\|_2^2 + \|g_i\|_2^2}{\bound^2}\right|> G\right\} \bigg| \siga\bigg)  \\
         &\quad \quad  =\pr\bigg(\bigcup_{i=k-\lambda + 1}^k \left\{\left|\frac{ \|\tilde g_i\|_2^2 - \|g_i\|_2^2 + 2 \|g_i\|_2^2}{\bound^2}\right|> G \right\}\bigg| \siga\bigg)\\
        &\label{bounddef1}\quad \quad \leq \pr\bigg(\bigcup_{i=k-\lambda + 1}^k \left\{\left|\frac{ \|\tilde g_i\|_2^2 - \|g_i\|_2^2}{\bound^2}\right| + 2> G\right\} \bigg| \siga\bigg)  \\
        & \quad \quad \leq \sum_{i=k-\lambda + 1}^k\pr\bigg(\left|\frac{ \|\tilde g_i\|_2^2 - \|g_i\|_2^2}{\bound^2}\right|> G - 2 \bigg| \siga\bigg)\\
        &\label{lchernoff1} \quad \quad \leq 2\lambda \exp \left(\frac{t^2 }{2Cp} - t\left(G - 2\right)\right),
    \end{align}
    where \cref{bounddef1} comes from \cref{gtoM},
    \cref{lchernoff1} comes from the \cref{chernoff-bound}.  Since \cref{lchernoff1} holds when $0\leq t \leq \frac{1}{\omega}$. We again must consider two cases to get the Chernoff bound. If the problem were unconstrained the minimum would occur at $t = Cp (G-2)$. Since this problem is constrained we first consider the case when  $Cp (G-2)\leq \frac{1}{\omega}$ at this point the Chernoff bound is 
    \begin{equation}
        2\lambda \exp\left(-\frac{Cp(G-2)^2}{2}\right).
    \end{equation}
    The second case occurs when $Cp (G-2) > \frac{1}{\omega}$ and in this case the function is minimized by setting $t = \frac{1}{\omega}$ at which point the bound is
    \begin{equation}
        2\lambda \exp\left(-\frac{(G-2)}{2\omega}\right). 
    \end{equation}
    Combing these two cases gives us that 
    \begin{align}
        \pr\bigg(&\bigcup_{i=k-\lambda + 1}^k \left\{\left|\frac{ \|\tilde g_i\|_2^2 + \|g_i\|_2^2}{\bound^2}\right|> G\right\} \bigg| \siga\bigg) \\
        &\leq 2\lambda \exp\left(-\min\left(\frac{Cp(G-2)^2}{2},\frac{(G-2)}{2\omega}\right)\right).
    \end{align}
    With the left and right terms of \cref{int-bound1} we can now progress to find the $G$ that minimizes \cref{int-bound1}. 
    \begin{align}
        \pr&\left(\left|\frac{\tilde \iota_k^\lambda - \iota_k^\lambda}{\bound^4}\right| > \epsilon\bigg| \siga\right) \\
        &\leq \inf_{G > 2} 2\exp\left(-\min\left( \frac{\epsilon^2 Cp\lambda}{2G^2 (1+\log(\lambda))},\frac{\epsilon\lambda}{2G\omega}\right)\right)\\ 
        &\quad \quad \quad +  2\lambda \exp\left(-\min\left(\frac{Cp(G-2)^2}{2},\frac{(G-2)}{2\omega}\right)\right). \label{final-rate-iota1}
    \end{align}
    We can then observe that when $G \geq 2 + \sqrt{\epsilon \lambda} $ it is the case that
    \begin{equation}
        \begin{aligned}
        \exp&\left(-\min\left(\frac{Cp(G-2)^2}{2},\frac{(G-2)}{2\omega}\right) \right) \\
        & \leq \exp\left(-\min\left( \frac{\epsilon^2 Cp\lambda}{2G^2 (1+\log(\lambda))},\frac{\epsilon\lambda}{2G\omega}\right)\right).
        \end{aligned} 
    \end{equation}
    We can upper bound the right-hand term of \cref{final-rate-iota1} in the following manner,
    \begin{align}
        &\inf_{G > 2} 2\exp\left(-\min\left( \frac{\epsilon^2 Cp\lambda}{2G^2 (1+\log(\lambda))},\frac{\epsilon\lambda}{2G\omega}\right)\right) \\
        & \quad \quad \quad \quad \quad+  2\lambda \exp\left(-\min\left(\frac{Cp(G-2)^2}{2},\frac{(G-2)}{2\omega}\right)\right)\\
        &\leq \inf_{G > 2 + \sqrt{\epsilon \lambda}} 2(1+\lambda)\exp\left(-\min\left( \frac{\epsilon^2 Cp\lambda}{2G^2 (1+\log(\lambda))},\frac{\epsilon\lambda}{2G\omega}\right)\right) \label{rchernff_min}\\
        &\leq 2(1+\lambda)\exp\left(-\min\left( \frac{\epsilon^2 Cp\lambda}{2(2 + \sqrt{\epsilon \lambda})^2 (1+\log(\lambda))},\frac{\epsilon\lambda}{2(2 + \sqrt{\epsilon \lambda})\omega}\right)\right).
    \end{align}
    Where the last line comes from recognizing that \cref{rchernff_min} is an increasing function of $G$. 
\end{proof}

\section{Proof of constant relative error}\label{rel:err}
We restate \cref{lemm:relative-error-M-iota} and provide a proof of the lemma below.
\begin{lemma} \label{lemm:relative-error-M-iota1}
    Under the conditions of \cref{thm:convergence-of-moments}, for $\epsilon > 0$, $\bound^4$ as described in \cref{sub-Gaussian:estimator}, $\tilde \iota_k^\lambda$ as defined in \cref{tildiota}, 
    \begin{equation}
    \begin{aligned}
    &\mathbb{P}\left( \left| \frac{\bound^4 - \tilde \iota_k^\lambda}{\bound^4}\right| > 1 + \epsilon, \bound^4 \neq 0  \bigg| \siga \right)\\ 
    &\leq 2(1+\lambda)\exp\Bigg(-\min\Bigg(\frac{\epsilon^2 Cp \lambda }{2(2 + \sqrt{\lambda \epsilon})^2 (1+ \log(\lambda))},
    \frac{\lambda \epsilon}{2 (2 + \sqrt{\lambda \epsilon})\omega}\Bigg)\Bigg).
    \end{aligned}
    \end{equation}
\end{lemma}
\begin{proof}
First,
\begin{equation} \label{eq:relative-error-decomposition}
\begin{aligned}
\left| \frac{\bound^4 - \tilde \iota_k^\lambda}{\bound^4}\right| &
\leq \left| \frac{\bound^4 - \iota_k^\lambda}{\bound^4}\right| 
+  \left| \frac{\iota_k^\lambda - \tilde \iota_k^\lambda}{ \bound^4} \right|,
\end{aligned}
\end{equation}
Moreover,
\begin{equation}
    \iota_k^\lambda \in \left[ \sigma_{\min}(A^{\top}B^{1/2})^4 \Vert \mathcal{P}B^{1/2}r_k  \Vert_2^4, \bound^4 \right].
\end{equation}
Applying this fact to \cref{eq:relative-error-decomposition},
\begin{equation}\label{fact-bound1}
\begin{aligned}
\left| \frac{\bound^4 - \tilde \iota_k^\lambda}{\bound^4}\right| &
\leq 1 
+ \left| \frac{\iota_k^\lambda - \tilde \iota_k^\lambda}{\bound^4} \right|.
\end{aligned}
\end{equation}
We now apply a relative error version of the bound in \cref{thm:cons-of-iota},\footnote{For a detailed derivation of relative error bound see \cref{lem:rel-of-iota}.} which supplies 

\begin{equation}\label{no-relat1}
    \begin{aligned}
    &\mathbb{P}\left( \left| \frac{\iota_k^\lambda - \tilde \iota_k^\lambda}{\bound^4} \right| > \epsilon \bigg{|} \siga \right) \\
    & \leq 2(1+\lambda)\exp\Bigg(-\min\Bigg(\frac{\epsilon^2 Cp \lambda }{2(2 + \sqrt{\lambda \epsilon})^2 (1+ \log(\lambda))},
    \frac{\lambda \epsilon}{2 (2 + \sqrt{\lambda \epsilon})\omega}\Bigg)\Bigg).
\end{aligned}
\end{equation}
The result follows by combining \cref{fact-bound1} and \cref{no-relat1}.
\end{proof}

\section{Shallow Water Model} \label{sec:SW}

For the experiment in \cref{subsec:4dvar}, we use the one dimensional shallow water problem as defined in work of Dimet et al. \cite{DimetVar}, which involves solving the following system of partial differential equations:
\begin{equation}\label{diffeq:phi}
    \frac{\partial \phi(x,t)}{\partial t} = - \frac{\partial}{\partial x} (\phi(x,t) u(x,t))
\end{equation}
and
\begin{equation}\label{diffeq:u}
    \frac{\partial u(x,t)}{\partial t} = - \frac{\partial}{\partial x} \left(\phi(x,t) + \frac{u(x,t)^2}{2}\right). 
\end{equation}
The functions $\phi(x,t)$ and $u(x,t)$ are unknown functions of the position, $x$, and time point, $t$. The function $\phi(x,t)$ represents the potential energy at a location at a particular time, and $u(x,t)$ represents the velocity at a location at a particular time. 

In this section we lay out the specifics of our 4D-Var problem by first discussing our simulation using Euler's method, then writing out our Jacobian equations used in incremental 4D-Var, before for finally presenting our modified version of \cref{algo:Adaptive-window}.
\paragraph{The Forward Model} To generate noisy observations to be assimilated, it is necessary to simulate the system. This simulation will be generated using a forward Euler method, which requires the discretization of the differential equations. In order to discretize the system we use $\Delta_t$ to represent an increment in time and $\Delta_x$ to indicate an increment in position. With this notation defined, the discretization of \cref{diffeq:phi} is 
\begin{align}
    \frac{\partial \phi(x,t)}{\partial t} \approx &\frac{\phi(x,t+ \Delta_t) - \phi(x,t)}{\Delta_t}= \\ \notag  & u(x,t) \frac{\phi(x - \Delta_x, t) - \phi(x + \Delta_x,t) }{2 \Delta_x} + \phi(x,t) \frac{u(x - \Delta_x, t)- u(x + \Delta_x,t)}{2 \Delta_x}, 
\end{align}
and the discretization of \cref{diffeq:u} is 
\begin{align}
    \frac{\partial u(x,t)}{\partial t} \approx &\frac{u(x,t+ \Delta_t) - u(x,t)}{\Delta_t} = \\ \notag &\frac{\phi(x - \Delta_x, t) - \phi(x + \Delta_x,t)}{2 \Delta_x} + u(x,t) \frac{u(x - \Delta_x, t) - u(x + \Delta_x,t)}{2 \Delta_x}.
\end{align}
From the discretization, we can then derive the state at a new time point for $\phi$ and $u$. Specifically, they are
\begin{align}
    &\phi(x,t+ \Delta_t) = \phi(x,t)  + \\ \notag &\quad \quad \quad   \Delta_t  \left(u(x,t)\frac{\phi(x - \Delta_x, t) - \phi(x + \Delta_x,t)}{2 \Delta_x} + \phi(x,t) \frac{u(x - \Delta_x, t) - u(x + \Delta_x,t)}{2 \Delta_x}\right), 
\end{align} 
for $\phi$ and
\begin{align}
    &u(x,t+ \Delta_t) = u(x,t)  + \\\notag &\quad \quad \quad \Delta_t \left( \frac{\phi(x - \Delta_x, t) - \phi(x + \Delta_x,t)}{2 \Delta_x} + u(x,t) \frac{u(x - \Delta_x, t) - u(x + \Delta_x,t)}{2 \Delta_x}\right),
\end{align}
for $u$.

\paragraph{The tangent model}\label{par:tangent} For incremental 4D-Var it is necessary to not only have the forward model, but also the first order linearization of that model \cite{talagrand1987variational}. This linearization requires the calculation of the Jacobian of the discretized model in terms of the functions $u$ and $\phi$ at all possible values of $x$ \cite{kalnay_2002}.  The nonzero Jacobian values at a particular position $x$ are:
\begin{equation}
    \frac{\partial \phi(x,t + \Delta_t)}{\partial \phi(x + \Delta_x, t)}  = - \frac{\Delta_t}{2 \Delta_x}u(x,t), 
\end{equation}
\begin{equation}
    \frac{\partial \phi(x,t + \Delta_t)}{\partial \phi(x, t)}  = 1 + \Delta_t \frac{u(x - \Delta_x, t) - u(x + \Delta_x,t)}{2 \Delta_x}, 
\end{equation}
\begin{equation}
    \frac{\partial \phi(x,t + \Delta_t)}{\partial \phi(x - \Delta_x, t)}  = \frac{\Delta_t}{2 \Delta_x} u(x,t), 
\end{equation}
\begin{equation}
    \frac{\partial \phi(x,t + \Delta_t)}{\partial u(x + \Delta_x, t)}  = -\frac{\Delta_t}{2 \Delta_x} \phi(x,t), 
\end{equation}
\begin{equation}
    \frac{\partial \phi(x,t + \Delta_t)}{\partial u(x, t)}  =  \Delta_t \frac{ \phi(x - \Delta_x, t) - \phi(x + \Delta_x,t)}{2 \Delta_x},
\end{equation}
\begin{equation}
    \frac{\partial \phi(x,t + \Delta_t)}{\partial u(x - \Delta_x, t)}  = \frac{\Delta_t}{2 \Delta_x} \phi(x,t), 
\end{equation}

\begin{equation}
    \frac{\partial u(x,t + \Delta_t)}{\partial \phi(x + \Delta_x, t)}  = - \frac{\Delta_t}{2 \Delta_x}, 
\end{equation}
\begin{equation}
    \frac{\partial u(x,t + \Delta_t)}{\partial \phi(x, t)}  = 0, 
\end{equation}
\begin{equation}
    \frac{\partial u(x,t + \Delta_t)}{\partial \phi(x - \Delta_x, t)}  = \frac{\Delta_t}{2 \Delta_x},
\end{equation}
\begin{equation}
    \frac{\partial u(x,t + \Delta_t)}{\partial u(x + \Delta_x, t)}  = - \frac{\Delta_t}{2 \Delta_x} u(x,t),
\end{equation}
\begin{equation}
    \frac{\partial u(x,t + \Delta_t)}{\partial u(x, t)}  = 1 + \Delta_t \frac{u(x - \Delta_x, t)- u(x + \Delta_x,t)}{2 \Delta_x},
\end{equation}
and
\begin{equation}
    \frac{\partial u(x,t + \Delta_t)}{\partial u(x - \Delta_x, t)}  = \frac{\Delta_t}{2 \Delta_x} u(x,t). 
\end{equation}

\subsection{A Reduced Memory Algorithm for 4D-Var}

When solving the Incremental 4D-Variational data assimilation problem, we wish to find the incremental update $u^k$ to an initial state estimate $z^{k-1}$ by solving
\begin{equation}
    \min_{u^{k}} \frac{1}{2}\left(\|(z^{k-1} - z_b) - u^{k}\|_V^2 + \sum_{i = 0}^{N_t} \|H^{k-1}_iM^{k-1}_{0,i}u^{k} - (y_i - H_i(x_i^{k-1}))\|_W^2\right).
\end{equation}
Here $V$ is the inverse covariance matrix for the background states, $W$ is the inverse covariance matrix for the observations, $N_t$ is the number of time points observed in the data, $z_b$ is the background state, $z^{k-1}$ is the current state estimate, $y_i$ is the observation at the $i^\text{th}$ time point, $x^{k-1}$ is the result of forward Euler applied to $z^{k-1}$ from time $0$ to time $i$, $H^{k-1}_i$ is the Jacobian of the observation function, $H_i$, and $M^{k-1}_{0,i}$ is the product of Jacobian matrices of the dynamics from time point zero to time point $i$ \cite{gurol:hal-03224132}. Under this setup, this problem can be solved using least squares solvers, and has a convenient structure that can be exploited by a row solver such that the memory load is minimized. The main structural advantage comes from us being able to generate $M^{k-1}_{0,i}$ as needed as we progress through the algorithm, meaning that by running the algorithm in a row-wise fashion  and let $N_c$ be the number of state variables, we only need to store a matrix of dimension $N_c \times N_c$ rather than $N_t N_c \times N_c$. We can additionally use right sketching to reduce the number of columns, which further reduces the storage costs to $N_c \times p$, where $p$ is the sketch size.  This substantial reduction in memory cost allows our solver to avoid substantial slowdowns from the full matrix memory accesses encounter by Krylov based methods, such as LSQR. 

Due to the inherent structure of the matrix system in the 4D-Var problem and our choice of \cite{MillerGentl} as a solver, which can work on row based blocks of a matrix; we propose a modified version of \cref{algo:Adaptive-window}, that does not require access to the full matrix system in the 4D-Var problem. 

\begin{algorithm}[hp]
    
    \caption{Tracking and Stopping for Least Squares}
    \begin{algorithmic}[2]
    \REQUIRE{Random sketching method satisfying \cref{def:JL} and sketch size $s$.}
    \REQUIRE{$V \in \mathbb{R}^{N_c \times N_c}$, $W \in \mathbb{R}^{N_c \times N_c}$ $b \in \mathbb{R}^m$, $z_0 \in \mathbb{R}^{N_c}$, $z_b \in \mathbb{R}^{N_c}$,$u_0 \in \mathbb{R}^{s}$, $\{y_i : i\in \{1,\dots,N_t\}\}$.}
    \REQUIRE{Moving window size $\lambda_1 \leq \lambda_2 \in \mathbb{N}$.}
    \REQUIRE{$\alpha > 0, \xi_I > 0 , \xi_{II} > 0 , \delta_I \in (0,1)$, $\delta_{II} > 1$, $\eta \geq 1$, $\upsilon > 0$.}
    \REQUIRE{Function $\textbf{h}(.)$ which applies the observation operator to a vector.}
    \REQUIRE{Function $\textbf{ForwardEuler}(.)$ progress a state vector forward one time point.}
    \REQUIRE{Function $\textbf{Jacobian}(.)$ generates the Jacobian matrix at a particular time point based on the state vector.} 
    \STATE $k \leftarrow 0$, $k' \leftarrow \infty$, $\tilde \rho_0^{*} \leftarrow 0$, $\tilde \iota_{0}^{*} \leftarrow 0$, $\lambda \leftarrow 1$, $u_k \leftarrow\{0\}^{n}$, $R\leftarrow \{0\}^{s \times s}$,$Tab\leftarrow \{0\}^{s }$, $D\leftarrow \{0\}^{s }$. 
    \WHILE{$k == 0$ \OR $\tilde \rho_{k-1}^\lambda \geq \upsilon$ \OR 
        \begin{align} \sqrt{\tilde \iota_{k-1}^\lambda} \geq \min\Bigg\{&\frac{\lambda (1-\delta_{I})^2\upsilon^2 Cp}{(1 + \log(\lambda))2\log(1/\xi_{I})\sqrt{\tilde \iota_{k-1}^\lambda}}, \frac{\lambda \upsilon (1-\delta_{I})}{2\log(1/\xi_{I}) \omega} ,\notag \\ &\frac{\lambda (\delta_{II}-1)^2\upsilon^2 Cp}{(1 + \log(\lambda))2\log(1/\xi_{II}) \sqrt{\tilde \iota_{k-1}^\lambda}}, \frac{\lambda \upsilon (\delta_{II} - 1)}{2\log(1/\xi_{II}) \omega} \Bigg\} \nonumber \end{align}}
    \STATE \# Iteration $k$ \#
    \STATE Generate $S_k$ and $(AS)_0 \leftarrow S_k$.
    \STATE $r_0 \leftarrow u_k - z_b + z_0$
    \STATE $\tilde g_k \leftarrow V (AS)_0^\top r_0$ 
    \STATE Use Gentleman's,\cite{MillerGentl}, on the problem $\min_{u_p}\|(AS)_0 u_p - r\|_{V}^2$, to update $R$, $Tab$, and$D$.
    \STATE Set $evol_0 \leftarrow z_0$
    \FOR{i = 1:ntimes}
    \STATE $evol_i \leftarrow \textbf{ForwardEuler}(evol_{i-1})$.
    \STATE $M_i \leftarrow \textbf{Jacobian}(evol_i)$.
    \STATE $(AS)_i \leftarrow M_i (AS)_{i-1}$. 
    \STATE $r_k = M_i u_k - (y_i - \textbf{h}(evol_i))$
    \STATE $\tilde g_k \leftarrow \tilde g_{k-1} + W (AS)_{i-1}^\top r_k$ 
    \STATE Use Gentleman's,\cite{MillerGentl}, on $\min_{u_p}\|(AS)_k u_p - r_k\|_{W}^2$, to update $R,Tab,\text{and}D$.
    \ENDFOR
    \STATE $u_{k+1} \leftarrow R^{-1}Tab$ 
    \STATE Perform Lines \ref{update_eq_start} - \ref{Credible-Interval} of \cref{algo:Adaptive-window}.



    \STATE $x_{k+1} \leftarrow x_k - S_k u_{k+1}$
    \STATE $ D \leftarrow \{0\}^s$, $ Tab \leftarrow \{0\}^s$,$ R \leftarrow \{0\}^{s \times s}$
    \STATE $k \leftarrow k + 1$
    \ENDWHILE
    \RETURN{$x_{k}$ \AND estimated $(1-\alpha)$-interval}
    \end{algorithmic}
    \label{alg:4dvar}
\end{algorithm}

\section{Estimating the distributional constants}

For accurate intervals, it is important to estimate the constants $C$ and $\omega$ appearing in \cref{def:JL}. To estimate these constants we lay out the following simulation study to determine conservative values of these constants for each of the three sampling types (Achlio, Gauss, FJLT), when the sample size is at least two. The experiment proceeds in accordance with the following steps.

\paragraph{Constant Estimation}
\begin{enumerate}
    \item For a chosen sketching method of (Achlio, FJLT, or Gauss) generate a sketching matrix $S_i \in \mathbb{R}^{p \times 128}$ and a random vector $x_i \in \mathbb{R}^{128}$ with entries from a $\text{Uniform}(0,1)$ distribution. With these values compute 
    \begin{equation}
        E_i = \frac{|\|Sx\|_2^2 - \|x\|_2^2|}{\|x\|_2^2}
    \end{equation}
    repeat $10,000,000$ independently. 
    \item Use the relative errors from the previous step approximate the tail probability of the distribution by computing
    \begin{equation}
       \Pr(E_i > \delta) \approx P_\delta = \frac{1}{10,000,000}\sum_{i=1}^{10,000,000} (E_i > \delta)
    \end{equation}
    for $\delta$ on a one dimensional grid ranging from 1 to 20 by 0.01.
    \item Remove any delta values corresponding to a  $P_\delta < 5/10,000,000$.
    \item Find the largest $\delta$ and the corresponding $P_\delta$. 
    \item Compute $\omega = \frac{\delta}{2\log(2/P_{\delta})}$.
    \item Compute the variance of the relative errors, which can be referred to as $\text{Var}$ and set $C = p \text{Var}$.
    \item Repeat this experiment 50 times and choose the largest $C$ and $\omega$ from these 50 trials.
\end{enumerate}

\begin{remark}
    The choice to remove all the values of $\delta$ with an empirical probability of being exceeded less than $5/10,000,000$ arises from the density of observations being too sparse for us to believe that the empirical probabilities are good representations of the true probabilities.
\end{remark}
\begin{remark}
    The choice of an initial dimension of $128$ is somewhat arbitrary. In some preliminary experiments it seemed the initial dimension did not have a large impact on the estimation of these constants. Thus, we chose the relatively small dimension of $128$.
\end{remark}

The results for running this experiment with $p=2$ can be found in \cref{p2}. We present the maximum value observed, which is chosen to be the estimate for the constants, as well as the mean and variance of the constants observed from the 50 trials.

\begin{table}
    \centering
    \caption{Constants for $p = 2$}
    \begin{tabular}{lcccc} \toprule
        Sketch Type & Constant & Max & Mean & Variance\\\midrule
        Gauss& $C$ & 1.1 & 1.09 & $2 \times 10^{-5}$\\
        Gauss & $\omega$ & 0.47& 0.44& .0002\\
        Achlio & $C$ & 1.16 & 1.14 & $4\times 10^{-5}$\\
        Achlio & $\omega$ &0.46& 0.40& 0.0003\\
        FJLT & $C$ & 0.83 & 0.83 & $2\times 10^{-5}$\\
        FJLT & $\omega$ &0.70& 0.62& 0.0009\\ \bottomrule
    \end{tabular}
    \label{p2}
\end{table}

These values are still quite conservative as it does appear that increasing $p$ leads to tighter values for both $C$ and $\omega$ as can be seen in \cref{p5}. This indicates that in fact at larger embedding dimensions more exact constants can be used.

\begin{table}
    \centering
    \caption{Constants for $p = 5$}
    \begin{tabular}{lcccc} \toprule
        Sketch Type & Constant & Max & Mean & Variance\\\midrule
        Gauss& $C$ & 2.07 & 2.06 & $5 \times 10^{-5}$\\
        Gauss & $\omega$ & 0.23& 0.21& $4 \times 10^{-5}$\\
        Achlio & $C$ & 2.16 & 2.15 & $6\times 10^{-5}$\\
        Achlio & $\omega$ &0.22& 0.20& $5\times 10^{-5}$\\
        FJLT & $C$ & 1.62 & 1.60 & $3\times 10^{-5}$\\
        FJLT & $\omega$ &0.32& 0.29& 0.0002\\ \bottomrule
    \end{tabular}
    \label{p5}
\end{table}

\bibliographystyle{templain}
\bibliography{least_square}
\end{document}